\documentclass[sigconf]{acmart}

\usepackage{todonotes}
\usepackage{amsmath}
\usepackage{amsfonts}
\usepackage{algorithmic}
\usepackage{subfig}
\usepackage{balance}

\usepackage[ruled, vlined, linesnumbered]{algorithm2e}
\usepackage{multirow}
\usepackage{mathtools}

\newcommand{\etal}{\emph{et~al. }}
\newcommand{\eg}{\emph{e.g., }}
\newcommand{\ie}{\emph{i.e., }}
\DeclareMathOperator*{\argmax}{arg\,max} 
 
\newcommand{\nop}[1]{}

\AtBeginDocument{%
  }

\setcopyright{acmlicensed}
\copyrightyear{2018}
\acmYear{2018}
\acmDOI{XXXXXXX.XXXXXXX}

\acmJournal{JACM}
\acmVolume{37}
\acmNumber{4}
\acmArticle{111}
\acmMonth{8}

\begin{document}

\title{Forecasting Attacker Actions using Alert-driven Attack Graphs}
\author{Ion Băbălău}
\affiliation{%
  \institution{Delft University of Technology}
  \city{Delft}
  \country{Netherlands}
}
\email{i.babalau@student.tudelft.nl}

\author{Azqa Nadeem}
\affiliation{%
  \institution{University of Twente}
  \city{Enschede}
  \country{Netherlands}}
\email{a.nadeem@utwente.nl}

\renewcommand{\shortauthors}{Băbălău et al.}

\begin{abstract}
While intrusion detection systems form the first line-of-defense against cyberattacks, they often generate an overwhelming volume of alerts, leading to alert fatigue among security operations center (SOC) analysts. 
Alert-driven attack graphs (AGs) have been developed to reduce alert fatigue by automatically discovering attack paths in intrusion alerts. However, they only work in offline settings and cannot prioritize critical attack paths. 
This paper builds an action forecasting capability on top of the existing alert-driven AG framework for predicting the next likely attacker action given a sequence of observed actions, thus enabling analysts to prioritize non-trivial attack paths. We also modify the framework to build AGs in real time, as new alerts are triggered. 
This way, we convert alert-driven AGs into an early warning system that enables analysts to circumvent ongoing attacks and break the cyber killchain.
We propose an expectation maximization approach to forecast future actions in a reversed suffix-based probabilistic deterministic finite automaton (rSPDFA).
By utilizing three real-world intrusion and endpoint alert datasets, we empirically demonstrate that the best performing rSPDFA achieves an average top-3 accuracy of 67.27\%, which reflects a 57.17\% improvement over three baselines, on average. 
We also invite six SOC analysts to use the evolving AGs in two scenarios. Their responses suggest that the action forecasts help them prioritize critical incidents, while the evolving AGs enable them to choose countermeasures in real-time. 

\end{abstract}



\maketitle

\section{Introduction}
Alert fatigue is one of the most common issues faced by security operations center (SOC) analysts, caused by the excessive volume of alerts that need to be investigated \cite{fireeye}. 
Existing research has proposed a number of ways to reduce this alert volume, \eg using alert correlation to group alerts likely triggered by the same attacker action \cite{ning2002constructing,qin2004discovering,sadoddin2006alert,zhu2006alert,salah2013model,alserhani2016alert,wang2016alert,haas2018gac,shittu2015intrusion,mcelwee2017deep}. 
%
However, these approaches cannot reconstruct attack paths, requiring analysts to manually investigate multiple alerts to create an attack timeline.

\begin{figure*}[t]
    \centering
     \subfloat[Modifications proposed to the SAGE \cite{nadeem2021alert} workflow (denoted by dotted edges): As new alerts arrive, the model is relearned using all alerts (evolving AG support). The S-PDFA is reversed into the rSPDFA, which is used to predict the next likely attacker action for each partial path. The refreshed AGs thus contain predictions for every partial attack path.]{\includegraphics[width=0.6\linewidth]{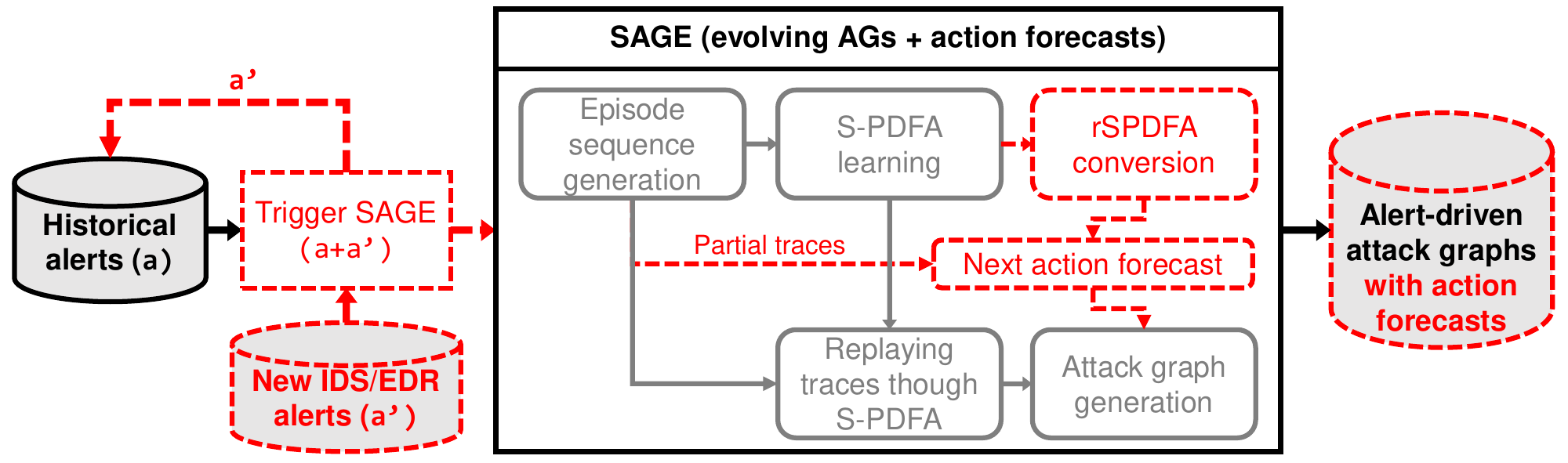}
    \label{fig:pipeline}}
    \hfil
\subfloat[An exemplary AG of attack stage (AS) and targeted service (TS) for victim1. The attack paths for attackers are shown in different edge colors. \underline{This paper:} the partial path (right) ends in action <AS, TS> with the probability $p$.]{\includegraphics[width=0.35\linewidth]{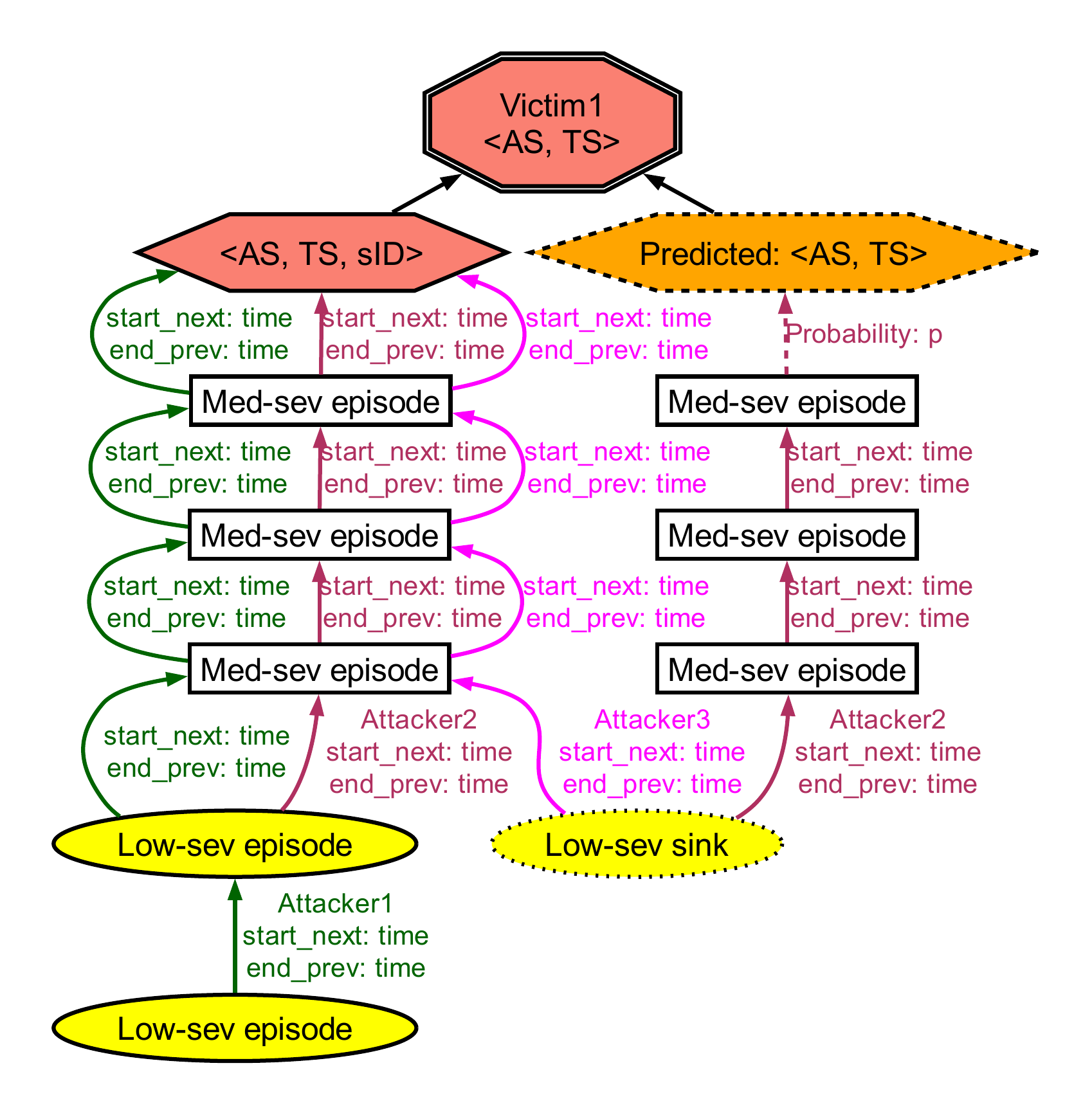}
    \label{back:ag}}    
    \caption{}
    \label{fig:overallpipeline}
\end{figure*}

Attack graphs (AG) are famous for modeling attack campaigns \cite{kaynar2016taxonomy} --- they visualize all possible pathways attackers can use to penetrate a network. 
%
Traditional AG generation approaches rely on network topology and vulnerability information \cite{noel2009advances}, which is often unavailable in operational settings. In contrast, Nadeem \etal have developed SAGE \cite{nadeem2021alert} -- the first machine learning based method to generate AGs without any expert input, using only intrusion alerts. 
Although SAGE aids in monitoring the current state of the network, it does so in an offline setting with no support for real-time responses. Moreover, SAGE's alert-driven AGs reflect only past attacker actions and lack future risk assessment -- SAGE fails to visualize partial attack paths (reflecting ongoing attacks), as it cannot predict their progression. This limitation poses challenges for monitoring active threats in operational settings. 

In this paper, we transform SAGE into an early warning system (EWS) capable of forecasting the next likely attacker action given a partial attack path. A \textit{partial path} is defined as an alert sequence ending in a low/medium severity alert. The forecasts allow SAGE to model ongoing attacks and help analysts prioritize critical paths. We modify SAGE to incrementally create updated AGs whenever new alerts are triggered. These enhancements provide support for real-time responses, and enable analysts to prioritize non-trivial attack paths, such as those beginning with low-severity alerts that could lead to critical exploits. 
%
We also show that SAGE can easily be extended to support multiple data sources utilized by SOCs, such as intrusion detection systems (IDS) and endpoint detection and response (EDR) systems.
The modifications are shown in Figure \ref{fig:pipeline}.

The sparsity of severe alerts remains a major hurdle in modeling attack paths and forecasting attacker actions \cite{nadeem2021alert}. SAGE addresses this by proposing a novel \textit{suffix-based} probabilistic deterministic finite automaton (S-PDFA) that exploits the structure of an attack campaign to highlight infrequent (severe) alerts, while also differentiating between contextually different attack campaigns. However, the suffix-model predicts the past based on the future, so it cannot directly be used to predict future actions. 
%
Thus, we propose to reverse the S-PDFA into rSPDFA, such that we can predict the future based on the observed actions. Since this reversal makes the model non-deterministic, we develop an expectation maximization algorithm that traverses the S-PDFA in reverse, keeps track of multiple possible futures in order to compute a probability distribution over the next likely actions, and returns the action with the highest probability. 

We evaluate our approach on three real-world alert datasets, of which two are collected in a commercial SOC, and one is collected through a student penetration testing competition. We quantitatively evaluate the performance of the forecasting module against three baselines by generating AGs in a streaming setting.
We also conduct semi-structured interviews with six security analysts from the commercial SOC to understand the efficacy of the evolving attack graphs for investigating ongoing incidents and determining remedial countermeasures. 
The results demonstrate that the proposed approach is the best at modeling the training data and generalizing to unseen test data compared to standard approaches. In fact, our forecasting modules obtain an average top-3 accuracy of 61.02\% with negligible runtime, while the baselines are only able to achieve an average accuracy of 42.8\%. The expert interviews show that the action forecasts help SOC analysts prioritize critical incidents, while the evolving AGs enable them to take remedial actions in real-time. 
Our main contributions are:


\begin{enumerate}
    \item We transform SAGE into an early warning system by developing a novel algorithm that enables us to forecast the next likely attacker action given a partial attack path. 
    \item We embed SAGE in operational settings by creating evolving AGs and supporting EDR logs (incl. intrusion alerts).
    \item We empirically evaluate the action forecasting module on alerts generated through a security testing competition.
    \item We conduct interviews with 6 SOC analysts to understand the efficacy of evolving AGs and the forecasting module. 
\end{enumerate}

\begin{figure*}[t]
    \centering
    \subfloat[S-PDFA]{\includegraphics[width=0.24\linewidth]{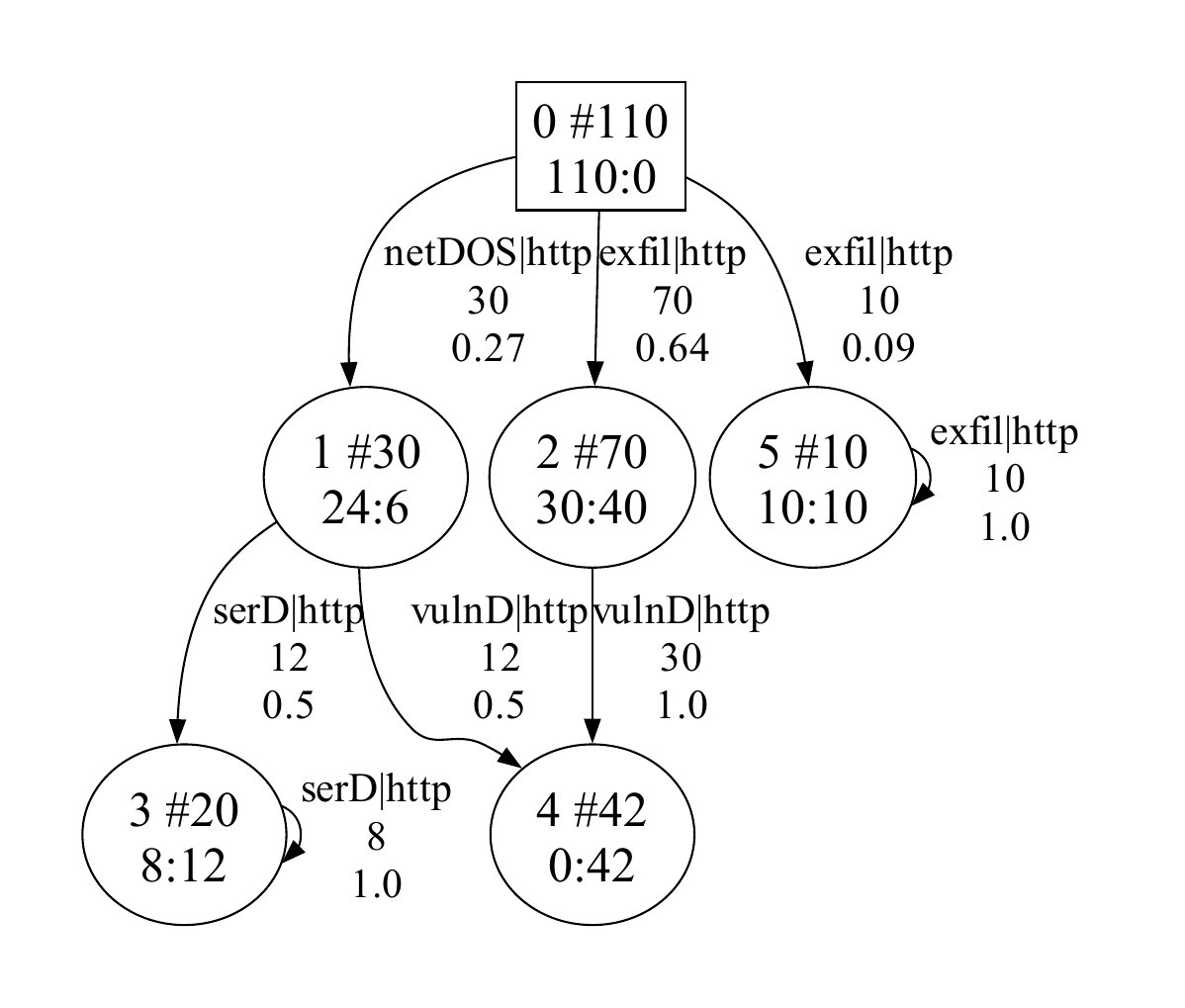}
    \label{fig:spdfa-example}}
    \subfloat[rSPDFA]{\includegraphics[width=0.25\linewidth]{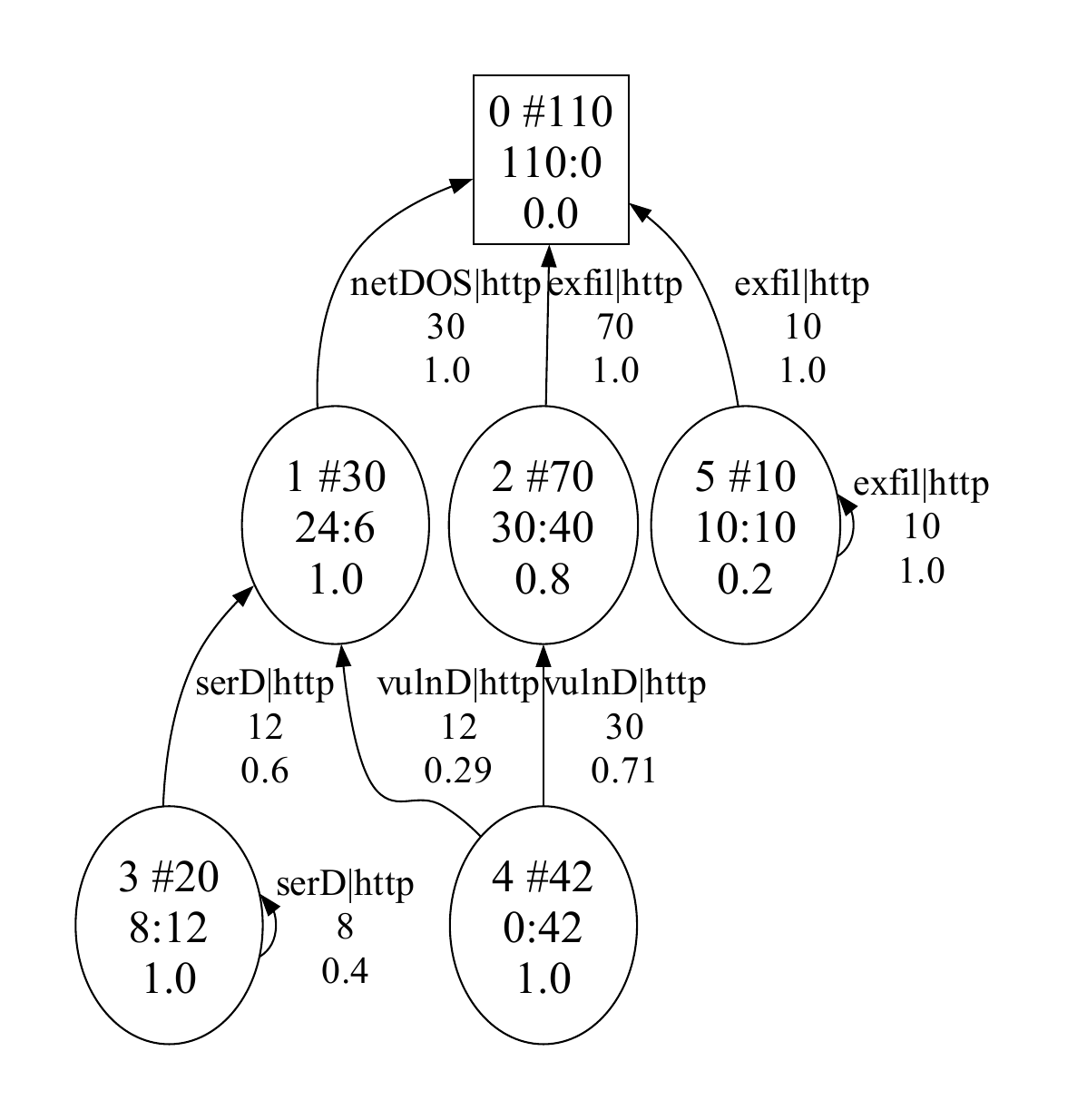}
    \label{fig:rspdfa-example}}
    \subfloat[PDFA]{\includegraphics[width=0.25\linewidth]{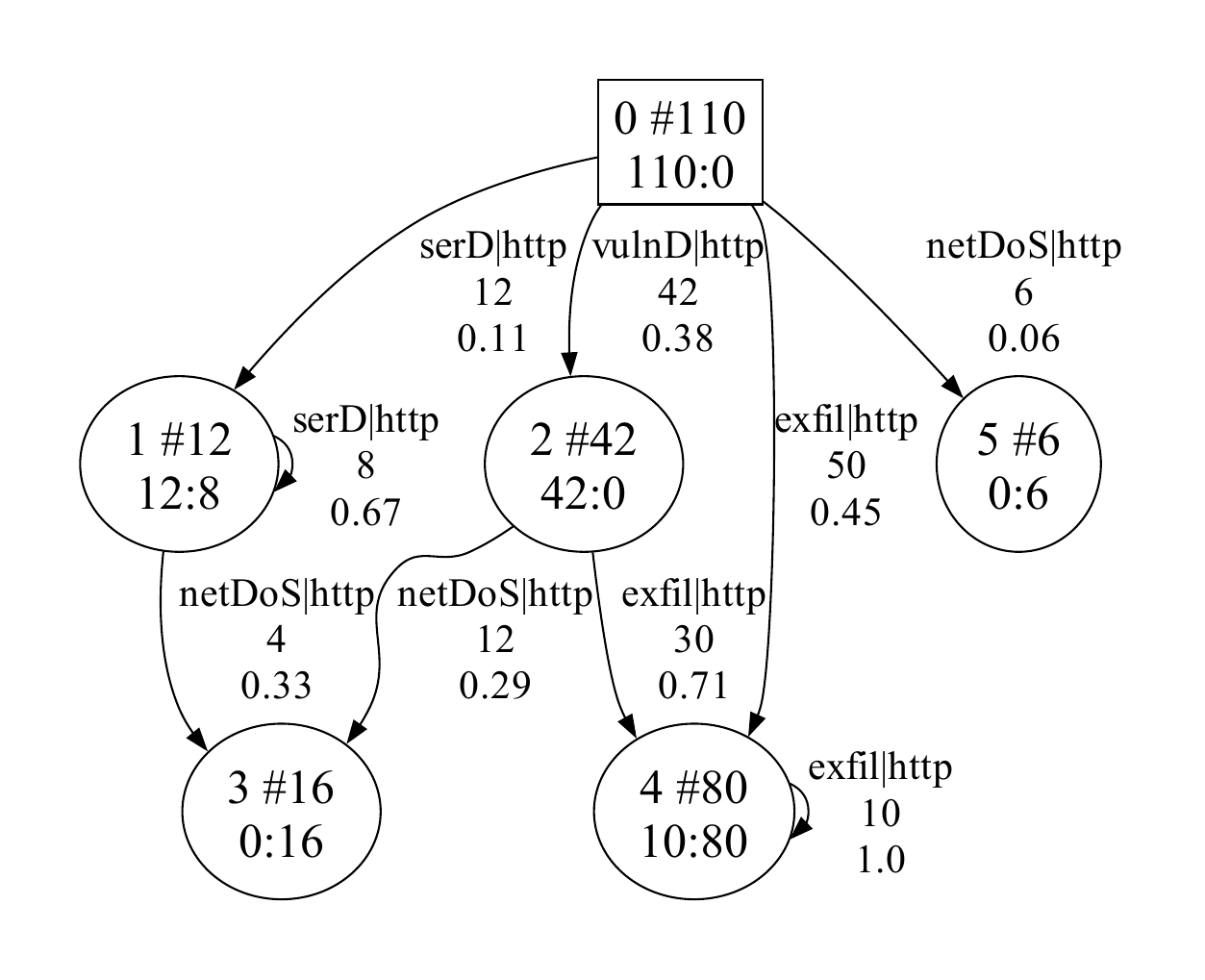}
    \label{fig:pdfa-example}}
  \subfloat[rSPDFA walkthrough]
    {\includegraphics[width=0.26\linewidth]{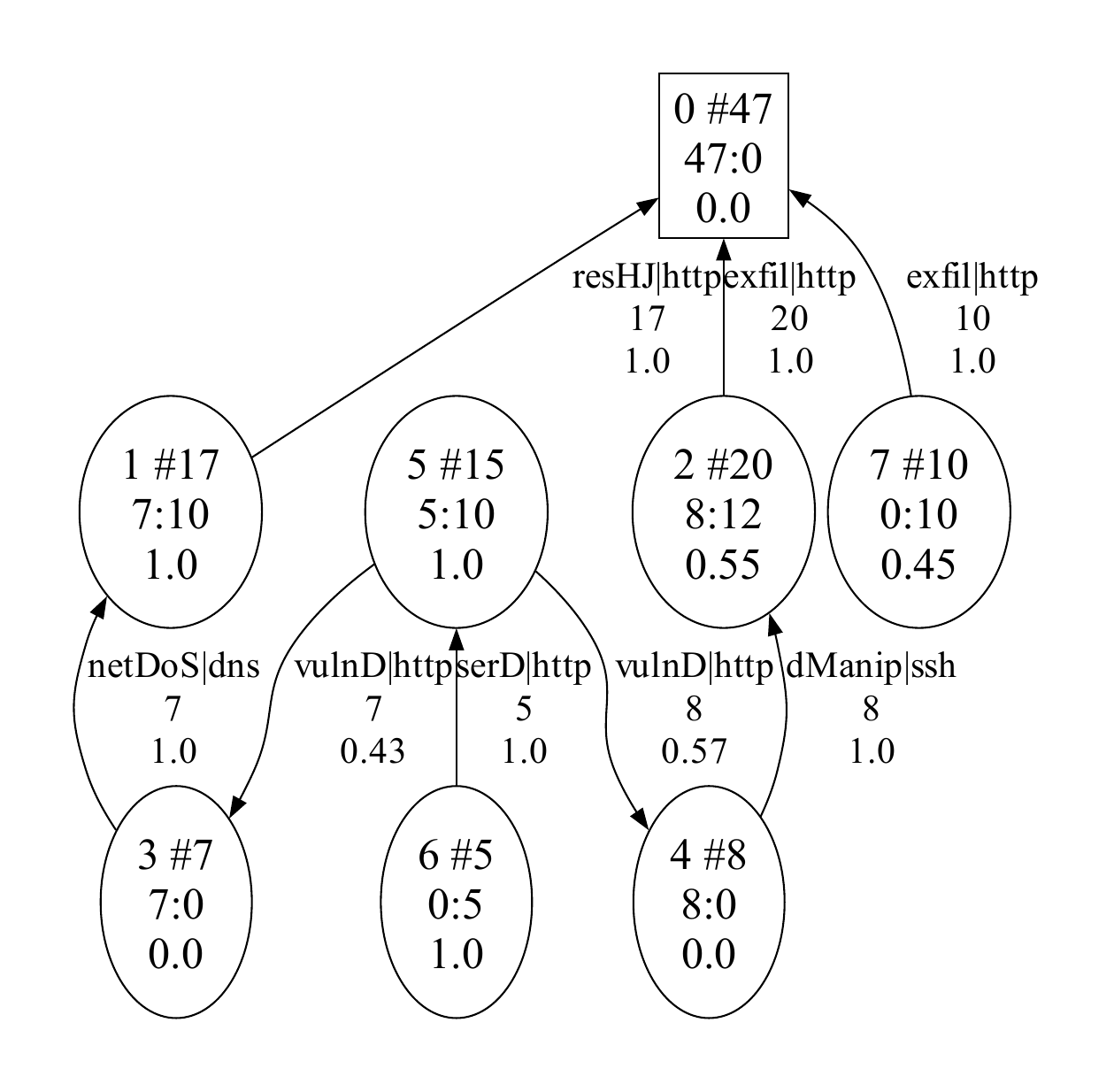}
    \label{fig:rspdfa-walkthrough}}   
\caption{(a) An exemplary S-PDFA. (b) Reversed rSPDFA from a. (c) PDFA learned from the same traces as a and b when they are not reversed. (d) An rSPDFA for demonstrating the abilities of traversal strategies FS, AS, HC.}
    \label{fig:model-examples}
\end{figure*}

\section{SAGE \& Alert-driven Attack Graphs}\label{sec:priliminary}

We start by providing background on SAGE and alert-driven AGs.

\paragraph{\textbf{SAGE workflow.}} SAGE\footnote{SAGE: \url{https://github.com/tudelft-cda-lab/SAGE}} (Figure \ref{fig:overallpipeline}) follows an interpretable sequence learning approach to create attack graphs (AG) without expert knowledge. It takes intrusion alerts as input, discovers the temporal and probabilistic relationships between alerts using an unsupervised suffix-based probabilistic deterministic finite automaton (S-PDFA), and represents them as attack campaigns in AGs. 


%
SAGE starts by aggregating alerts into episode sequences for each (source, destination) IP pair. The episodes are considered equivalent to attacker actions, and are characterized by a 4-tuple \textit{(start time, end time, attack stage, targeted service)}. Here, the attack stage maps alert signatures to the MITRE ATT\&CK framework via the Action-Intent framework (AIF) \cite{moskal2020cyberattack}, and the targeted service is derived from the most frequently occurring destination port in an episode. To create an episode, the frequency of the alerts is examined for a given attack stage. An increase in frequency denotes the start of an episode, while a decrease denotes the end (see \cite{nadeem2021alert}). The episodes in the sequences are time-ordered, usually beginning with low-severity episodes and ending in medium/high-severity episodes. 

The episode sequences are converted into traces to train the S-PDFA. 
An input trace contains univariate symbols derived from the episodes' \textit{(attack stage|targeted service)}, and is reversed (due to suffix model) such that the severe attack stages are towards the start of the trace. 
The Flexfringe tool \cite{verwer2017flexfringe} iteratively applies state merging on the input traces until a compact S-PDFA is learned. 

\paragraph{\textbf{S-PDFA syntax.}} The S-PDFA is formally defined as a 5-tuple $A = \langle Q, \sum, \Delta, P, q_0 \rangle$, where  Q is a finite set of states; \(\sum\) is a finite alphabet of symbols;  \(\Delta\)  is a finite set of transitions;  \( P : \Delta \rightarrow [0, 1] \) is the transition probability function, and \(q_0 \in Q\) is the final state (due to suffix model).
\nop{\begin{itemize}
    \item Q is a finite set of states
    \item \(\sum\) is a finite alphabet of symbols
    \item \(\Delta\)  is a finite set of transitions
    \item \( P : \Delta \rightarrow [0, 1] \) is the transition probability function
    \item \(q_0 \in Q\) is the final state (due to suffix model)
\end{itemize}}
A transition \(\delta \in \Delta\) in an S-PDFA is a tuple \(\langle q, q', a \rangle\) where \( q, q' \in Q \) are the target and source states, and \( a \in \sum \) is a symbol. P is a function such that \( \sum_{q,a} P(\langle q, q', a\rangle ) = 1\). Additionally, \(\Delta \) is such that for every \( q \in Q \) and \( a \in \sum \), there exists at most one \( \langle q, q', a\rangle \in \Delta \), making the model (suffix) deterministic \cite{nadeem2021alert}. 

In an S-PDFA, the states are represented by vertices and transitions are represented by edges (see Figure \ref{fig:spdfa-example}). 
The state label contains: (i) state identifier (sID), (ii) total occurrence count (computed as the sum of all incoming transitions, symbolizing the number of times the incoming symbol appears in the training traces in the given context), (iii) number of traces in the training set that continue to other states, and (iv) number of traces that end in this state. 
The transition label contains: (i) transition symbol, (ii) transition count, and (iii) transition probability.

The S-PDFA is an interpretable and deterministic graphical model of all attack campaigns present in an alert dataset.
It addresses alert sparsity by exploiting the structure of an attack campaign -- the infrequent severe alerts that lie at the start of (reversed) traces are accentuated by the suffix model, enabling clear differentiation between similar alerts leading to different outcomes. This is done by modeling the contextual meaning behind alerts in state identifiers (sID) -- similar alerts that have different futures or pasts are modeled using different states in the S-PDFA. Finally, the S-PDFA has Markovian property that ensures that the incoming transition symbols to a given state are unique, thus allowing us to easily interpret the states as \textit{milestones} achieved by the attackers.

\paragraph{\textbf{Attack graph syntax.}} Once the S-PDFA is learned, a separate AG is extracted from the S-PDFA for every attack objective achieved on the victim host(s). 
Figure \ref{back:ag} shows the anatomy of an alert-driven AG. It is an aggregated representation of relevant alerts, where each attack path originates from one of the starting (\ie yellow) vertices (representing episodes) and leads to the root (\ie objective) vertex. 
Every attacker that obtains the objective is shown using a different edge color. Multiple attack attempts are shown as separate attack paths. 
However, partial attack paths (that do not yet lead to an objective) are not shown since SAGE cannot predict how they would unfold.
%

The vertex shape represents the severity of attacker actions (\ie oval for low, rectangle for medium, and hexagon for high). 
The context of an episode is denoted by sID (from the S-PDFA). The episodes that occur too infrequently to be used in statistical computations (sink states) are shown as dotted vertex borders.
%
Furthermore, a list of alert signatures is displayed by hovering over each vertex so that analysts can link them to the relevant intrusion alerts.

\section{Extending Alert-driven Attack Graphs For Action Forecasting}\label{sec:method}

Figure \ref{fig:pipeline} shows the updates made to the SAGE workflow -- we reverse the S-PDFA in order to forecast the next action for each partial path, and display it in the AGs. We re-trigger this workflow each time new alerts arrive to create evolving AGs that show the latest situation. We also extend the parser for endpoint logs. We release our code: \url{https://github.com/ibabalau/SAGE}.

\subsection{Forecasting Next Attacker Actions}

The S-PDFA model is exceptionally effective for modeling highly imbalanced alert distributions and generalizing to unseen alerts compared to Markov chains and other approaches, as shown in \cite{nadeem2021alert}. However, due to the suffix-based nature of the model, it cannot directly be used to forecast future actions. In contrast, a standard prefix-based PDFA can easily predict future actions, but it is ineffective for modeling infrequent severe alerts \cite{nadeem2021alert}. 
Besides, considering that the original premise of the SAGE tool was interpretability, we opt for using a single model for context modeling and action forecasting to promote trust among practitioners \cite{nadeem2022sok}.


%
To this end, 
we perform forecasting with the S-PDFA by reversing its transitions for the prediction task, such that we now traverse it from the bottom to the top. We call this derived model variant the reversed S-PDFA (rSPDFA), see example in Figure \ref{fig:rspdfa-example} for the S-PDFA in Figure \ref{fig:spdfa-example}. 
%
%
An rSPDFA has different properties compared to the original S-PDFA. 
(i) It is non-deterministic, so we must keep track of multiple possible futures (paths) to compute a probability distribution over next actions. 
(ii) The rSPDFA states can have multiple outgoing transitions with the same symbol because of the S-PDFA's Markovian property.
(iii) There is no single root/starting state, instead, the states without outgoing transitions in the S-PDFA become new root states in the rSPDFA.
(iv) The reversal causes a change in the state and transition probabilities. 
(v) The rSPDFA state labels now show the number of traces that \textit{start} from a state, instead of the number of traces that end in it. It also lists the \textit{(starting)} probability, \ie the probability of a trace starting from a state. 

Next, we explain the process of recomputing the probabilities for the rSPDFA, discovering matching paths for a trace, and computing the probability distribution over the next actions.

%

\nop{\begin{figure*}[t]
    \centering
    \subfloat[]{\includegraphics[width=0.2\linewidth]{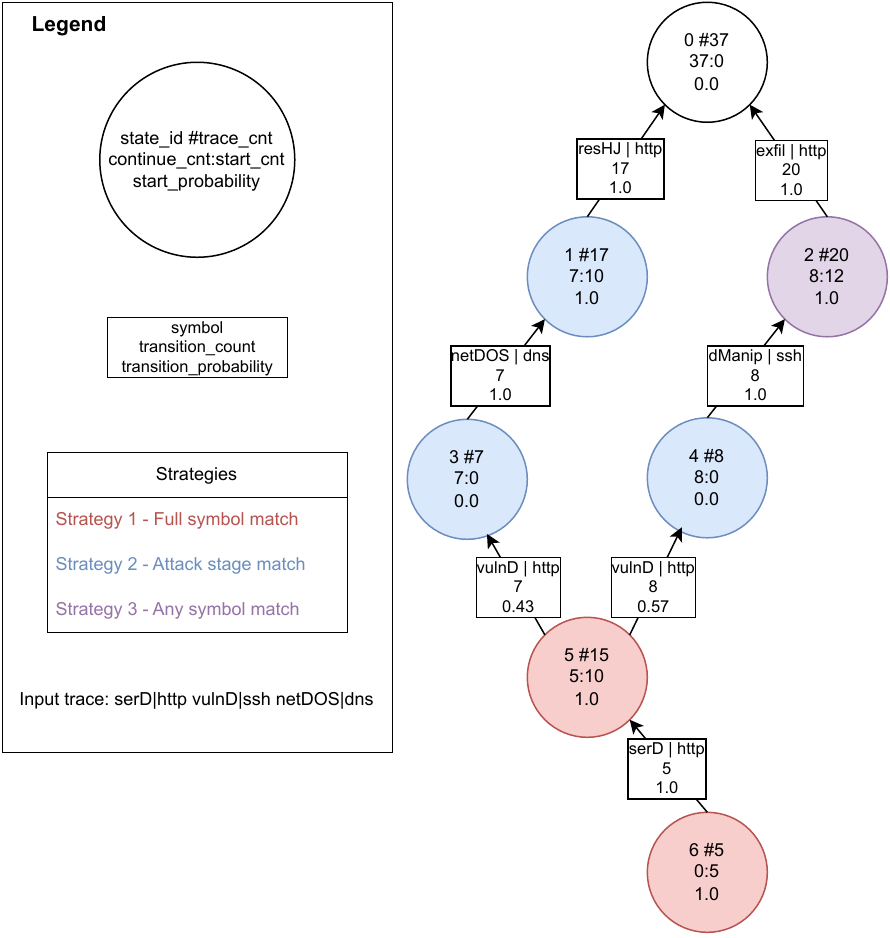}}
    \subfloat[]{\includegraphics[width=0.6\linewidth]{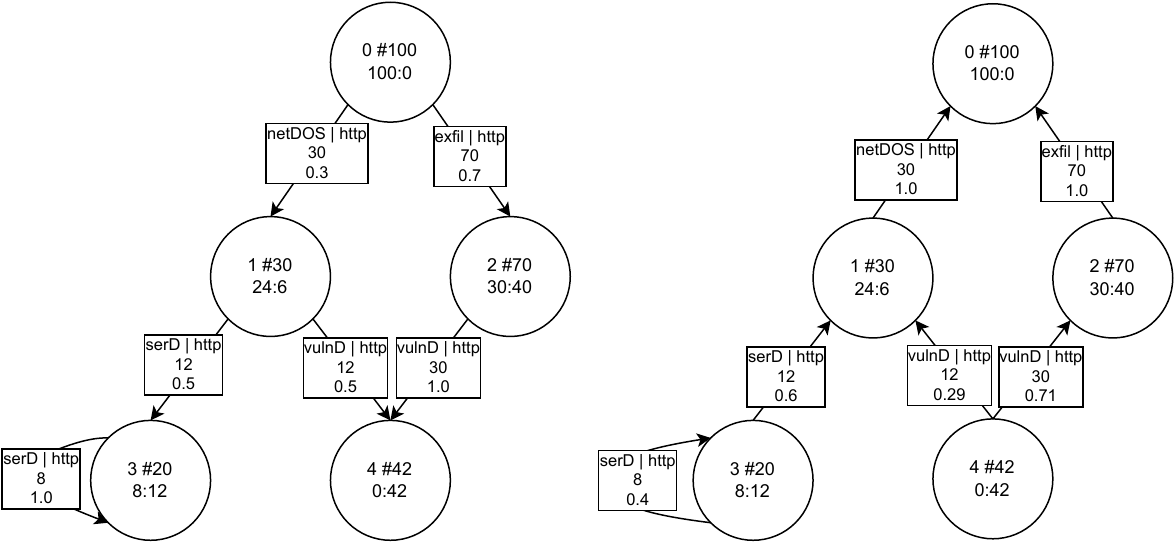}}
    \subfloat[]{\includegraphics[width=0.2\linewidth]{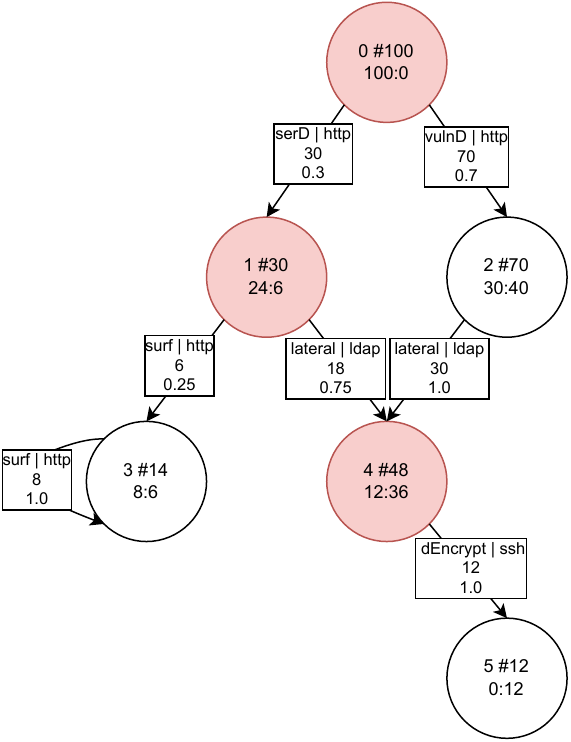}}
\caption{}
    \label{}
\end{figure*}}

\subsubsection{\textbf{Probability recalculation for rSPDFA}}
%
For a given transition $\langle d,s,a\rangle$ in the S-PDFA with states \( s,d \in Q \) and symbol \( a \in \sum \), the reversed transition probability $P(\langle s,d,a\rangle)$ in the rSPDFA is defined in Eq. \ref{eq:trans}.

\begin{equation}
P(\langle s,d,a\rangle) = \frac{count(\langle s,d,a\rangle)}{totalcount(d)}
\label{eq:trans}
\end{equation}

where \(count()\) returns the number of transitions from $d$ to $s$ on symbol $a$, and \(totalcount()\) returns the total occurrence count for the state $d$. For example in Figure \ref{fig:rspdfa-example}, the new transition probability from state 4 to state 2 with symbol \texttt{vulnD|http} is computed as: 

\[ P(\langle 2,4,\mathit{vulnD|http}\rangle) = \frac{count(\langle 2,4,\mathit{vulnD|http}\rangle)}{totalcount(4)} = \frac{30}{42} = 0.71\]

Every state has a starting probability. Since the rSPDFA inherits the Markovian property from the S-PDFA, it means that while a state \(s \in Q\) can have multiple outgoing transitions, they will all be associated to the same symbol \(a \in \sum\). Thus, the starting probability \(P_{a}(s)\) is computed using Eq. \ref{eq:start_state}.
\begin{equation}
P_{a}(s) = \frac{startcount(s)}{\sum_{s' \in S | \exists \langle ?,s',a \rangle} startcount(s')}
\label{eq:start_state}
\end{equation}

where \(startcount()\) returns the number of traces that start in $s$, and the set $S$ contains all the states with outgoing transition symbol $a$. For example, we compute the starting probability of state 2 as:

 \[ P_{\mathit{exfil|http}}(2) = \frac{startcount(2)}{startcount(2)+startcount(5)} = \frac{40}{40+10} = 0.8 \] 
 
 This means that there is an 80\% probability of a trace starting from state 2 given the first symbol \textit{exfil|http}.

\subsubsection{\textbf{Path finding algorithm}} \label{section:path_finding}
We search the rSPDFA for all paths that match a given trace of input symbols. We use the term \textit{path} to describe a sequence of states that can be reached by exploring transitions in the rSPDFA given a starting state and an input trace. 
Since the rSPDFA is non-deterministic, there will be more than one possible paths for a given trace. 
%
Further, we compute a \textit{reachable path} by traversing from one state to the next in the rSPDFA given the symbols in the trace until there are no more symbols left. 

We start by finding all states where the first symbol of the input trace occurs, forming a set of starting states $S$. Note that the starting states are not necessarily root nodes. Because the rSPDFA has Markovian properties (the previous state only depends on the next state, and not on the ones after it), the path finding algorithm can start from intermediary states as well.
We then adopt a depth-first search approach for finding reachable paths, \ie we recursively visit a state, examine the outgoing transitions that match the input symbol, and add the states associated to the matching transitions in the path. Once all the states are explored, we return the list of reachable paths. If we visit a state with no outgoing transitions while the input trace remains, we consider it to be an incomplete path and exclude it from the set of reachable paths. 

We implement memoization to speed up the path finding algorithm. We essentially use a look-up table where the key is (state identifier, input trace), and the value is the set of paths reachable from the state given the input trace. 

The decision to add a particular state to the path depends on the symbol matching criteria. We consider three traversal strategies:

\begin{enumerate}
    \item Strategy 1 - full symbol (FS): We visit a state if the input symbol matches the associated transition symbol (attack stage, targeted service).
    \item Strategy 2 - attack stage (AS): We visit a state if the input symbol matches the associated transition symbol (attack stage only).
    \item Strategy 3 - hybrid choice (HC): We first match via FS and AS. If no match is found, we visit the state with the highest associated transition count regardless of the input symbol.
\end{enumerate}

We consider these strategies to account for variations in the input traces that emerge from noisy alerting systems. For instance, HC can handle a trace where the attacker performs actions in a different order, while AS can handle cases where the targeted service changes. 

We use the rSPDFA in Figure \ref{fig:rspdfa-walkthrough} to exemplify the traversal strategies: 
Suppose we have observed three actions, \ie \textit{[serD|http, vulnD|ssh, netDOS|dns]}, and want to predict what will happen next. 
\textbf{Strategy 1:} For the first observed symbol \texttt{serD|http}, there is only one possible state to start from, \ie state 6, which leads to state 5. There is no match for the next input symbol \texttt{vulnD|ssh}. Thus, this strategy returns no reachable paths. 
\textbf{Strategy 2:} For the first observed symbol's attack stage \texttt{serD}, there is only one possible state to start from, \ie state 6, which leads to state 5. For the next symbol's attack stage \texttt{vulnD}, there are two possible states, \ie 3 and 4. The algorithm randomly picks state 3 to explore first. For the next symbol's attack stage \texttt{netDOS}, we reach state 1, which also symbolizes the end of the trace. Going back to explore state 4, there is no match for the next symbol's attack stage \texttt{netDOS}. Thus, this strategy results in one reachable path, \ie (6, 5, 3, 1).
\textbf{Strategy 3:} For the first observed symbol \texttt{serD|http}, there is only one possible state to start from, \ie state 6, which leads to state 5. For the next symbol \texttt{vulnD|ssh}, we match the attack stage, and obtain two possibilities, \ie state 3 and 4. Choosing state 3 for the next symbol \texttt{netDOS|dns}, we reach state 1, which also marks the end of the trace. Going back to state 4 and the symbol \texttt{netDOS|dns}, we pick the transition with the maximum count that leads us to state 2, marking the end of the trace. This strategy results in two reachable paths, \ie (6, 5, 3, 1), (6, 5, 4, 2).

Given a set of reachable paths $p \in paths$, we compute a probability distribution over the next possible actions, and return the one with the maximum likelihood. We achieve this by computing the probability of each reachable path $p = s_1\cdots s_N$, \ie multiplying the transition probabilities of each state, as well as the starting probability of the first state (to estimate the probability of a trace starting from that state), see Eq. \ref{eq:path_prob}.
For strategies AS and HC, because we explore paths which contain symbols that are different from the ones in the input trace (either in targeted service or the full symbol), we assign a weight correlated to the fraction of the symbol that matches so that we give a higher emphasis to the transitions where the full symbol matches. We multiply the transition probability with \textit{factor= f} if the attack stage matches, and with \textit{factor= 2*f} if the full symbol matches. We then normalize the probability so that the sum of path probabilities equals 1, \ie $\sum_{p \in paths} P(p) = 1$.

\begin{multline}
P(s_1\cdots s_n) =  \frac{P_{a_1}(s_1) * \prod_{i=1}^{n - 1} P(\langle s_{i+1},s_i, a_i\rangle) * \mathit{factor}}{\sum_{p \in paths} P(p)}
\label{eq:path_prob}
\end{multline}


Considering the two reachable paths returned by strategy 3, $\{p1, p2\} \in paths$, p1= (6, 5, 3, 1), p2= (6, 5, 4, 2), and assuming \textit{f=2}, we compute their path probabilities as:

\[
    P(p1) = \frac{\splitdfrac{P_{\mathit{serD|http}}(6) * P(\langle \mathit{6,5,serD|http} \rangle))*4*P(\langle 5,3,}{\mathit{vulnD|http}\rangle)*2*P(\langle 3,1,\mathit{netDOS|dns} \rangle)*4}}{18.32} = 0.75 
\]
\[
    P(p2) = \frac{\splitdfrac{P_{\mathit{serD|http}}(6) * P(\langle \mathit{6,5,serD|http} \rangle)*4*P(\langle 5,4,}{\mathit{vulnD|http} \rangle)*2*P(\langle \mathit{4,2,dManip|ssh}\rangle)*1}}{18.32} = 0.25
\]


\subsubsection{\textbf{Action forecasting}}
 
Given a set of reachable paths and their path probabilities, the probability distribution over the next action is computed as Eq. \ref{eq:nextprob}. The prediction result is the symbol with the maximum probability. 

\begin{equation}
    P(a) = \sum_{\mathclap{\substack{p \in paths\\ \textsc{NextAction}(p) = a}}}P(p)
    \label{eq:nextprob}
\end{equation}

We compute the probability distribution over the transition symbols occurring after the last visited state, \ie \texttt{resHJ|http} for p1 and \texttt{exfil|http} for p2. The probabilities of the next actions are: $P(resHJ|http) = P(p1) = 0.75$, and $P(exfil|http) = P(p2) = 0.25$. Thus, given the partial path \textit{[serD|http vulnD|ssh netDOS|dns]}, the next predicted symbol is \texttt{resHJ|http} with a probability of 75\%.

\subsection{Evolving Attack Graph Generation}
Since we can no longer assume that all alerts are available in an offline setting, we modify SAGE to generate attack graphs in a streaming setting. 
This is done by following an incremental approach: As soon as new alerts arrive, they are merged with the pool of historical alerts, and re-trigger the execution of SAGE. 
Because Flexfringe learns the S-PDFA in less than 0.5 seconds \cite{nadeem2021alert}, this approach can be used to re-trigger SAGE and regenerate the AGs as periodically as required by a SOC.

Each time the S-PDFA is relearned, the rSPDFA-based prediction module is applied to all partial paths present within the input traces, and the prediction result is used to determine the set of AGs in which these partial paths are visualized.
At each execution, new AGs are created for new victim hosts and novel attacks. Otherwise, old AGs are refreshed with newly observed attack paths since the last update.
Specifically, if we observe an attack path that ends in a high-severity action for which an AG already exists, we add the path to this AG. Similarly, if the prediction of a partial path is a high-severity action for which an AG already exists, we add the partial path and its prediction to this AG.
However, if an AG does not yet exist for this high-severity action, we create it with the newly observed attack paths (including the predictions for partial paths, if applicable).
We also create AGs for low- and medium-severity predictions with the predicted actions as the root nodes. These AGs contain partial paths from \textit{all relevant victim hosts} where this prediction was made. For example, for three hosts with the predicted action \textit{vulnerability discovery|HTTP}, we create a single AG with attack paths from these hosts. 
%
%
We visualize the predicted actions as orange vertices with dashed borders (see Figure \ref{back:ag}). The shape of the vertex reflects its severity. The edge leading to the prediction displays the prediction probability.

%
%
%
%

How often to trigger the re-execution of SAGE depends on the level of cyber-readiness required by a SOC. For instance, it can be executed frequently during a cyber attack, and a few times a day for regular monitoring. Moreover, the volume of historical alerts to include is also dependent on the relevance of the forecasts and the desired amount of context one wants to see in the AGs, \eg a sliding window over a specific time period vs. all historical alerts. A larger time window increases the average AG size, but allows analysts to get a broader view of activities carried out on a host.

\subsection{Adding Support for EDR Logs}
We extend the SAGE parser to support endpoint alerts, \eg from EDR systems, in addition to intrusion alerts. 
The modifications are based on the properties of the EDR alerts received from the commercial SOC. 
%
We observe the following differences between IDS and EDR alerts:
(a) EDR logs typically do not have IP and port information; 
(b) one alert may be generated for multiple hosts; 
(c) the alerts usually specify both MITRE tactic and technique as separate attributes; 
(d) the same attack stage appearing for multiple alerts may have different severity levels depending on the signature that was triggered, and 
(e) the volume of alerts is significantly sparser than the CPTC alert datasets used in \cite{nadeem2021alert}.

Our EDR log parser works as follows: 
%
%
(i) We start by splitting alerts that contain multiple host names, such that each alert is associated to a single host name keeping all other attributes the same. We also anonymize the host names for privacy reasons.
(ii) We specify the attack stage of each alert using both the MITRE tactic and technique. If an alert has both these attributes, the attack stage becomes \textit{<Tactic.Technique>}. For alerts with multiple tactics and/or techniques, we concatenate them in the form \textit{<Tactic1.Technique1, Tactic2.Technique1>} to prevent information loss. 
(iii) Due to alert sparsity, we convert each alert into an episode. The episode is then characterized by the attack stage of the enclosed alert, whereas the start and end times are the timestamps of the alert itself. No targeted service is specified since port information is not available.
(iv) Since source IP is often missing, the episode sequences are created on a per-host basis, instead of on a (source, destination) IP pair basis. This has virtually no impact on the resulting attack graphs, \ie the paths still show the sequence of events that led to an attack on a victim host.
(v) Finally, the input trace symbols for the S-PDFA are characterized by the episodes' \textit{(attack stage|severity)}, instead of \textit{(attack stage|targeted service)}. 
The input traces are then used as usual in the SAGE workflow (Figure \ref{fig:pipeline}). 

\section{Datasets And Experimental Setup}\label{sec:expsetup}

\begin{table}[t]
\caption{Summary of experimental datasets.}
\label{tab:dataset_all}
\resizebox{0.9\columnwidth}{!}{%
\begin{tabular}{lcccc}
\hline
\textbf{Dataset} & \textbf{Period} & \textbf{Alerts} & \textbf{Traces} & \textbf{Avg. seq length} \\ \toprule
\textbf{CPTC-2018} & 10 hours & 331554 & 555 & 5.78 \\ 
\textbf{Pentest} & 3 weeks & 136 & 24 & 3.5 \\ 
\textbf{Big env.} & 3 months & 529 & 83 & 4.4 \\ \bottomrule
\end{tabular}}
\end{table}

\begin{figure}[t]
    \centering
    \subfloat[CPTC-2018]{\includegraphics[width=0.5\linewidth]{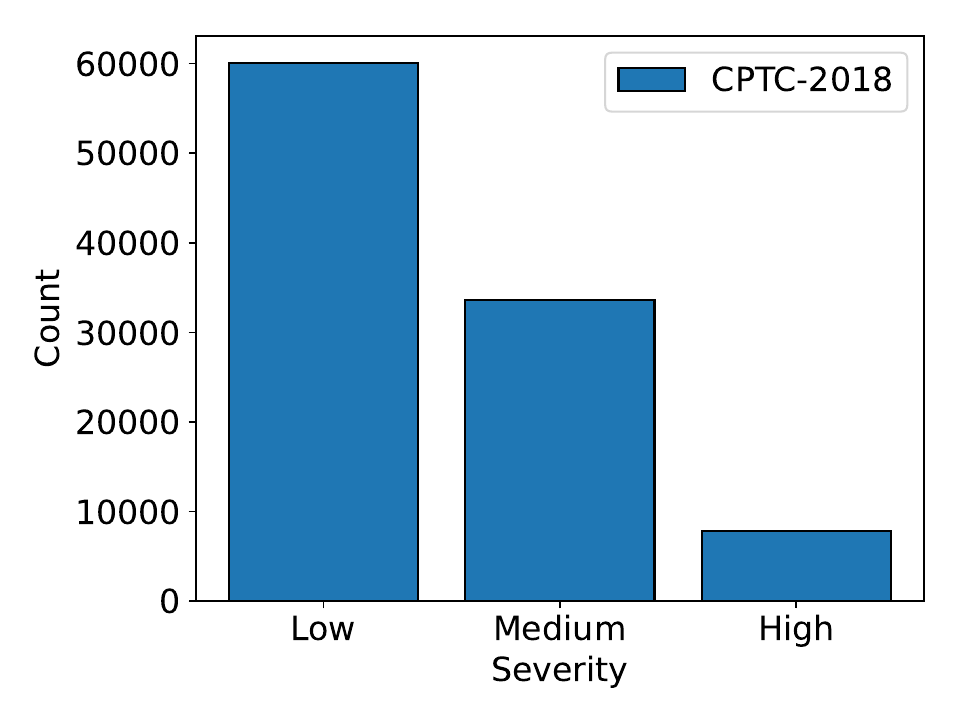}}
    \subfloat[Commercial SOC]{\includegraphics[width=0.5\linewidth]{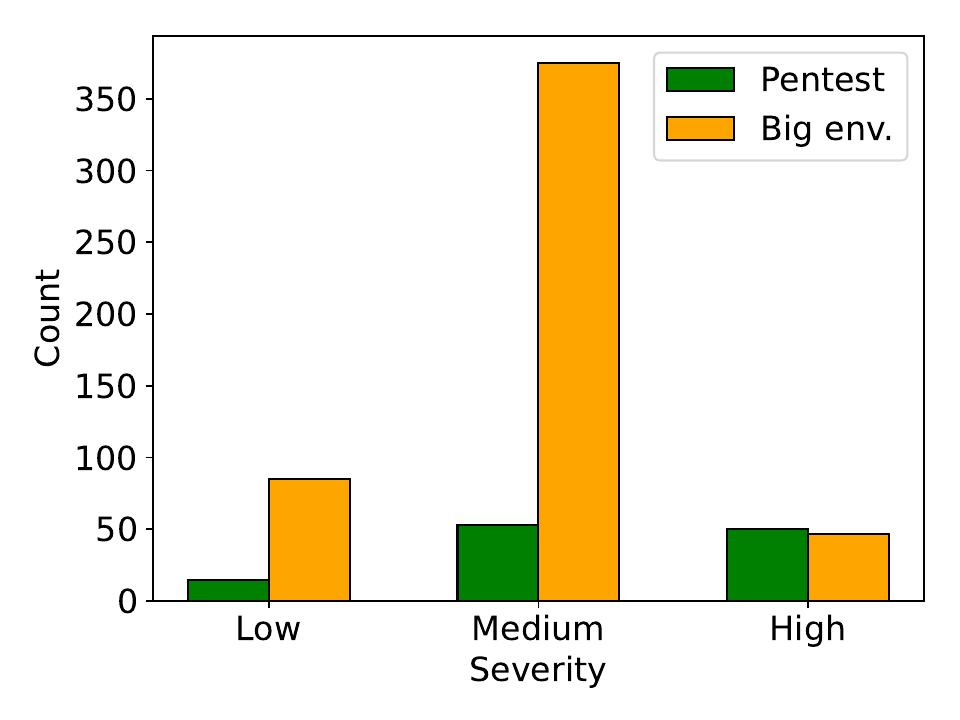}}
\caption{The alert severity distribution in the experimental datasets. High-severity alerts are the most infrequent alerts. }
    \label{fig:sevdistro-alldatasets}
\end{figure}

\noindent\paragraph{\textbf{Datasets}} We use three alert datasets for our experiments: one containing alerts from a student penetration testing competition, and two containing alerts from a commercial SOC to evaluate the performance of the action forecasting module and the ability of SAGE to generate attack graphs in real-time. A summary of the datasets is given in Table \ref{tab:dataset_all}.

\noindent\textbf{1) CPTC-2018:} The Collegiate Penetration Testing Competition (CPTC) challenges college students to demonstrate their penetration testing and security assessment skills \cite{cptc-2018-dataset}.
The CPTC-2018 contains alerts from six teams (T1, T2, T5, T7-T9) who were tasked to compromise a common fictitious automotive company's network. Each team was given access to fixed-IP machines that were used to conduct the attacks. A Suricata IDS monitored the traffic and collected alerts on a per-team basis. The competition lasted for 10 hours, resulting in 331,554 alerts. We extract the following attributes from the alerts: timestamp, alert signature, and the source and destination IP addresses and port numbers. Other than the attackers' IP addresses, no other ground truth is available for CPTC-2018.  
We obtain 555 traces (384 unique traces, 148 unique symbols) from the 331,554 alerts, where 61.9\% of the traces have lengths between 3 and 6 (see appendix Fig. \ref{pred:lendistrib}). The dataset also follows a long-tailed distribution in terms of severity, where the majority of the traces end in a low-severity symbol, likely symbolizing ongoing  attempts (see appendix Fig. \ref{fig:sevdistrib}). The alert imbalance in CPTC-2018 thus makes it a challenging dataset to be used for prediction tasks.

\noindent\textbf{2) Commercial SOC datasets:} We worked with Northwave Cybersecurity who provided access to alerts triggered by their Sentinel, NIDS, and EDR systems. The alerts contain attributes, such as timestamp, alert signature, alert severity, MITRE tactic and technique, and the entity that the alert refers to, \eg host name or IP address. 

We derive two alert collections: (1) \textbf{Pentest:} We collected alerts over a period of three weeks during a penetration test. 
The 136 alerts allow us to model the actions of a pen tester. (2) \textbf{Big Environment:} We collected alerts over a period of three months. The 529 alerts allow us to model suspicious behavior in a more `noisy' environment. 
Figure \ref{fig:sevdistro-alldatasets} shows the alert severity distribution across the three datasets. We observe that they have vastly different volumes and severity distributions. 
For Pentest, medium and high alerts are prevalent versus for CPTC-2018 (also a penetration test), low and medium are prevalent, capturing a variety of (novice and expert) penetration testing strategies. 

\noindent\paragraph{\textbf{Experiments}} We conduct three experiments to evaluate our proposed method using diverse IDS and EDR alerts: 
\nop{\begin{enumerate}
    \item We utilize CPTC-2018 to empirically measure the performance of the action forecasting module against three baselines.
    \item We utilize CPTC-2018, Pentest, and Big Environment alerts for testing AG evolution and alert triaging.
    \item We organize semi-structured interviews with six security analysts from Northwave to evaluate the usability of the evolving AG and the action forecasting module.
\end{enumerate}}

\noindent\textbf{1) Forecasting: } We utilize CPTC-2018 to empirically measure the performance of the action forecasting module against 3 baselines.

\noindent\textbf{2) Evolving AGs: } We utilize CPTC-2018, Pentest, and Big Environment alerts for testing AG evolution and alert triaging.

\noindent\textbf{3) User study: } We organize semi-structured interviews with 6 security analysts from Northwave to evaluate the usability of the evolving AG and the action forecasting module.

\noindent\paragraph{\textbf{Baselines}}
We compare the performance of the action forecasting traversal strategies (FS, AS, HC) against three baselines:

\noindent\textbf{1) Random guess:} The first naive baseline assigns a random probability mass to the set of available symbols, \ie $\frac{1}{\mathit{|\sum|}}$. Thus, for a given symbol $a_t$, the next symbol $\mathit{a_{t+1}}$ is chosen at random.

\noindent\textbf{2) Frequency based:} The second naive baseline computes the probability mass of a symbol based on the frequency of bigram occurrences. Specifically, we examine all ($a_t, a_{t+1}$) tuples in the training set. The probability mass assigned to $a_{t+1}$ is its normalized frequency of occurrence. For a given symbol $a_t$, we pick $a_{t+1}$ with the maximum probability of occurrence.

\noindent\textbf{3) PDFA based:} The natural competitor to an automaton-based prediction method is a standard prefix-based PDFA, which predicts the future based on observed instances. We train a PDFA using the same parameters and training data as the S-PDFA (without reversing the traces, see example in Figure \ref{fig:pdfa-example}). The resulting PDFA is deterministic, so only a single reachable path is possible for every input trace. We always start at the root state, and traverse the states based on the symbols in the input trace. If a matching transition for an input symbol cannot be found, we pick the transition with the highest count to account for noisy traces (similar to strategy 3). We do this until the end of the input trace is reached. The symbol with the maximum transition count is given as the prediction result with a probability of $\frac{\mathit{count(\langle s,d,a \rangle)}}{\mathit{\sum_i count(\langle s,d,i \rangle)}}$. If we reach a state with no outgoing transitions, then we consider it an incomplete path. 


%
%
%

\noindent\paragraph{\textbf{Parameters}} As a default, we set up SAGE to re-execute in hourly intervals, and provide predictions for all partial paths in the datasets. Because the rSPDFA's non-deterministic nature makes path exploration computationally expensive (particularly for AS and HC), we consider at most $t$ symbols in a partial path for the prediction task for all experimental variants. Thus, we use the observations from $[0 \cdots t]$ to predict the action at time $t+1$, and use the actual observation at time $t+1$ as ground truth. 
We set this threshold to $t=5$ as the upper-bound, such that a maximum of 5 symbols are used to predict the $6^{th}$ symbol. Shorter traces are included as well. 
This threshold was set based on the mean length of the traces and runtime (see appendix Figures \ref{pred:lendistrib} and \ref{fig:testing_time}).
To avoid the risk of over-fitting in sparse datasets, we employ k-fold cross validation to evaluate the action forecasting module. 
SAGE takes k-1 chunks of input traces to train the S-PDFA and rSPDFA, and uses the last k chunk of input traces for the prediction task.
We set $k=5$ as the best trade-off between accuracy and fit (see appendix Figure \ref{pred:kfold}).
For the multiplication factor of symbol matching, we set $f=55$ based on the best accuracy (see appendix Figure \ref{fig:factor}).
For training the automaton models, we set $\mathit{state\_count,\  symbol\_count,\ sink\_count}$ to 5 based on \cite{nadeem2021alert}.

\noindent\paragraph{\textbf{Evaluation metrics}}
The prediction performance of the various methods is measured using the following metrics:

\noindent\textbf{1) Perplexity:} It measures the prediction power of a model, \ie how well it fits the training data vs. how well it generalizes to unseen test data. It is defined as $2^{-\frac{1}{N}\sum_{i=1}^N log_2P(x_i)}$, where \textit{N} is the number of traces and \textit{P(x)} is the trace probability. Lower values are better. Perplexity was used in \cite{nadeem2021alert} to measure model quality.  


\noindent\textbf{2) Top-3 AS accuracy:} The top-3\footnote{We report top-3 accuracy to account for imbalance in alert severities.} attack stage (AS) accuracy reports the fraction of true labels where the attack stage matches a symbol in the top-3 predicted actions. We report this accuracy for low, medium, and high severity attack stages individually. 

\noindent\textbf{3) Top-3 UTAS accuracy:} The top-3 unseen trace attack stage (UTAS) accuracy reports the generalizability of the prediction module to unseen traces, \ie the fraction of true labels where the attack stage matches a symbol in the top-3 predicted actions of a previously unseen test set. For this, we create a test set with traces that were not present in the training data, neither as individual traces, nor as part of another longer trace (\textit{N}=19).

\noindent\textbf{4) No path found rate:} The fraction of traces for which the path finding algorithm returned no reachable path.

\noindent\textbf{5) Runtime:} The time it takes for the prediction module to predict the next action for each trace on average.

\noindent\paragraph{\textbf{Analyst interviews}}
For the third experiment, we organized semi-structured interviews with six SOC analysts from Northwave.
\textit{While it is a small sample size, it reflects roughly the median number of participants in security user studies, as reported by \cite{nadeem2022sok}}.
We present two scenarios to the analysts and pose questions regarding a) the usefulness of evolving AGs for obtaining updated situational awareness, b) the usefulness of the action forecasting module for selecting effective countermeasures, and c) the comparison of alert-driven AGs with currently used alert management tools. We follow the think-aloud protocol to understand their thought processes when interpreting the AGs.

\begin{table}[t]
\caption{Participant demographics for the interviews.}
\label{tab:demographics}
\resizebox{0.5\columnwidth}{!}{
\begin{tabular}{ccc}
\hline
 & \textbf{Designation} & \textbf{Experience} \\ \toprule
\textbf{P1} & Tier-3 analyst & 4-5 years \\ 
\textbf{P2} & Tier-2 analyst & 2-3 years \\ 
\textbf{P3} & Tier-2 analyst & 1-2 years \\ 
\textbf{P4} & Tier-2 analyst & 1-2 years \\ 
\textbf{P5} & Tier-2 analyst & 1-2 years \\ 
\textbf{P6} & Tier-2 analyst & 2-3 years \\ \bottomrule
\end{tabular}}
\end{table}

\noindent\textbf{Recruitment:} The interviews were conducted with one tier-3 analyst and five tier-2 analysts. They had 1-5 years of experience, see Table \ref{tab:demographics}. The interviews were conducted using Microsoft Teams, where each interview lasted 30 minutes. We start by giving each participant a brief crash-course on alert-driven AGs. We then ask questions from a prepared script, and record the answers to transcribe them later. Following an inductive thematic analysis, we report on common themes that appear in these transcriptions. We obtained the necessary IRB approvals from the university's ethics board for the user study.

\noindent\textbf{Study design:} We ask the participants to use alert-driven AGs in two hypothetical scenarios. 
In the first scenario, the participants are shown an AG generated from the Pentest alerts, and are asked to interpret it. Then, they are shown an updated version of the AG (generated from additional alerts) and asked the same question again.
In the second scenario, the participants are shown an AG generated from CPTC-2018, and are asked for their recommended remedial steps. Then, they are shown the prediction(s) for the attack path(s), and are asked whether their recommendation changes.
In the last part of the interview, the participants are asked to compare alert-driven AGs with the alert management tools they currently employ for incident response.

\section{Results and Discussion}
In this section, we report on the results of our three experiments.
%

\subsection{Action forecasting performance}

\begin{figure}[t]
\centering
\includegraphics[width=0.75\linewidth]{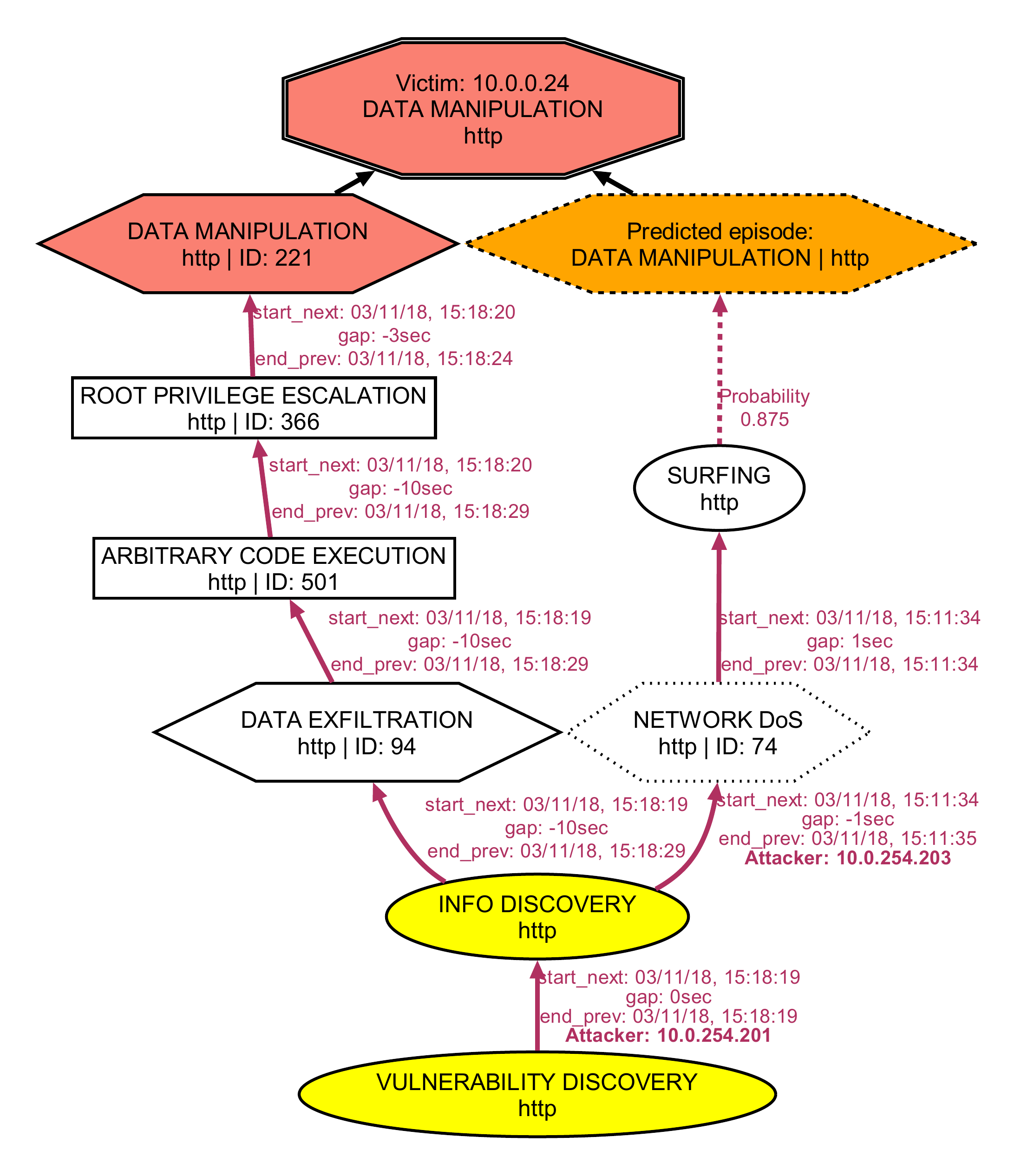}
\caption{Data manipulation conducted over http for victim 10.0.0.24 showing one observed path and one partial path with an 87.5\% probability of ending in data manipulation.}
\label{fig:predictionAG}
\end{figure}

\begin{table*}[t]
\caption{Forecasting performance of our proposed algorithm against baselines. The best values are \textbf{bold}, and the second best values are \textit{italicized*}. Strategy 3 performs the best while strategy 2 achieves the best trade-off between runtime and performance.}
\label{tab:resultsmain}
\resizebox{2\columnwidth}{!}{%
\begin{tabular}{lcccccccccc}
\hline
\textbf{Method} & \multicolumn{2}{c}{\begin{tabular}[c]{@{}c@{}}\textbf{Perplexity}\end{tabular}} & 
\multicolumn{3}{c}{\begin{tabular}[c]{@{}c@{}}\textbf{Top-3 AS accuracy}\end{tabular}} & \multicolumn{1}{c}{\begin{tabular}[c]{@{}c@{}}\textbf{Top-3} \\ \textbf{UTAS accuracy}\end{tabular}} & \multicolumn{1}{c}{\begin{tabular}[c]{@{}c@{}}\textbf{Avg. accuracy} \\ \textbf{AS+UTAS}\end{tabular}} & \multicolumn{1}{c}{\begin{tabular}[c]{@{}c@{}}\textbf{No path} \\ \textbf{found rate}\end{tabular}} & {\begin{tabular}[c]{@{}c@{}}\textbf{Runtime} \\ \textbf{(sec)}\end{tabular}} \\ \cline{2-3} \cline{4-6} 
 & \multicolumn{1}{c}{\textbf{Train}} & \multicolumn{1}{c}{\textbf{Test}} &  \multicolumn{1}{c}{\textbf{Low}} & \multicolumn{1}{c}{\textbf{Medium}} &  \multicolumn{1}{c}{\textbf{High}} &  &  &  \\ \midrule
\textbf{Random guess} & - & - & 
\multicolumn{1}{c}{59.29} & \multicolumn{1}{c}{2.77} & 8.33 & 37.08 & 26.87 & -  & - \\ 
\textbf{Frequency based} & - & - & 
\multicolumn{1}{c}{85.54} & \multicolumn{1}{c}{18.51} & 2.77 & 50.00 & 39.21 & - & - \\
\textbf{PDFA} & \multirow{1}{*}{51092.02} & \multirow{1}{*}{67283.08} & 
\multicolumn{1}{c}{86.72} & \multicolumn{1}{c}{48.14} &  \textit{57.40*} & 57.05 & 62.33 & \textit{0.014*} & \textbf{4.37 x $\mathbf{10^{-6}}$} \\ 
\textbf{rSPDFA (FS strat 1)} & \multirow{3}{*}{\textbf{49986.25}} & \multirow{3}{*}{\textbf{55458.02}} & 
\multicolumn{1}{c}{72.27} & \multicolumn{1}{c}{44.44} & 47.22 &  37.91  & 50.46 &  0.230  & \textit{1.55 x $\mathit{10^{-4}}$*} \\ 
\textbf{rSPDFA (AS strat 2)} & & & 
\multicolumn{1}{c}{\textit{87.61*}} & \multicolumn{1}{c}{\textit{50.92*}} & \textit{57.40*} & \textit{65.41*} & \textit{65.34*} & 0.064 & 6.55 x $10^{-3}$ \\ 
\textbf{rSPDFA (HC strat 3)} & & & 
\multicolumn{1}{c}{\textbf{88.20}} & \multicolumn{1}{c}{\textbf{54.62}} & \textbf{58.33} & \textbf{67.91} & \textbf{67.27} & \textbf{0.003} & 1.17 x $10^{-2}$ \\  \bottomrule
\end{tabular}}
\end{table*}

Figure \ref{fig:predictionAG} shows an AG generated through our forecasting module from CPTC-2018 for the victim 10.0.0.24. We have observed one attack path starting from vulnerability discovery that ended in data manipulation. The dataset also contained a partial path that started in info discovery and ended in surfing. For the offline SAGE, it was unclear where to display this partial path. Through our forecasting module, we predict that the next likely action for this path is also data manipulation with a probability of 87.5\%, and thus we decide to place it in this AG.

Table \ref{tab:resultsmain} shows the performance of our three forecasting modules against three baselines on CPTC-2018. These results essentially represent a multi-class classification problem with 148 outcomes (symbols) on a highly imbalanced dataset, highlighting the difficulty of the prediction task.  
We first compute Perplexity on the automaton models by using an 80-20 split. The models are less ``perplexed'' on the training data vs. the test data, as expected. However, between the standard PDFA and our suffix-based rSPDFA, the latter consistently achieves a better Perplexity value for both training and test data. This suggests that the S-PDFA approach is superior at modeling attack campaigns and generalizing to unseen patterns (as was also reported in \cite{nadeem2021alert}).

%
The top-3 AS accuracy shows the prediction performance of the variants on different alert severities. We observe that the HC and AS strategies outperform other approaches for all low, medium, and high severity predictions, and that there exists a non-linear relationship between alert frequency and accuracy. The naive baselines struggle with accurate predictions, especially for the infrequent medium and high severity alerts. Surprisingly, the PDFA achieves a comparable accuracy to the (second-best) AS strategy for high-severity predictions, despite them being the most infrequent alert type.
However, in terms of generalizability measured by the top-3 UTAS accuracy, the HC and AS strategies again outperform all baselines with a significant margin. Particularly, the PDFA is substantially worse (\ie 10.86 percentage points worse) than the rSPDFA. 
Moreover, in terms of the three rSPDFA strategies,  FS strategy consistently under-performs compared to AS and HC because it is too restrictive. Especially in an incremental learning set up where the targeted service can change, this strategy will find no match just because the service did not match. In essence, it provides a conservative lower-bound for the rSPDFA performance, which is still significantly better than the naive baselines, on average. 

Runtime is an important factor when selecting a forecasting module. The PDFA is naturally the fastest approach for action forecasting (because of determinism). For the non-deterministic models, the more paths they explore, the longer runtime we observe. Thus, it makes sense why the HC strategy achieves the highest runtime among the rSPDFA models. Having said that, it is also important to consider whether these algorithms manage to find any reachable paths at all. The results show that the HC strategy discovers at least one reachable path for 99.997\% cases since it is partially a greedy approach, followed by the PDFA model (which is also greedy). AS achieves the most number of paths among non-greedy approaches. 
%
%

Combining these results altogether, we observe that the rSPDFA approach is the best suited for action forecasting on highly imbalanced datasets -- it not only outperforms existing baselines in terms of prediction accuracy and generalizability to previously unseen attacks, but we can also utilize it to effectively model the contextual meaning behind alerts and improving the actionability of the AGs. 
Among the rSPDFA traversal strategies, the AS and HC strategies are the best options. HC and AS achieve an average accuracy of 67.27\% and 65.34\%, respectively (versus 62.33\% for the best performing baseline, PDFA). Among the two, we conclude that AS achieves the best trade-off between prediction accuracy and runtime.

\begin{figure*}[t]
    \centering
    \subfloat[Scenario 1a (Evolving attack graphs)]{\includegraphics[width=0.3\linewidth]{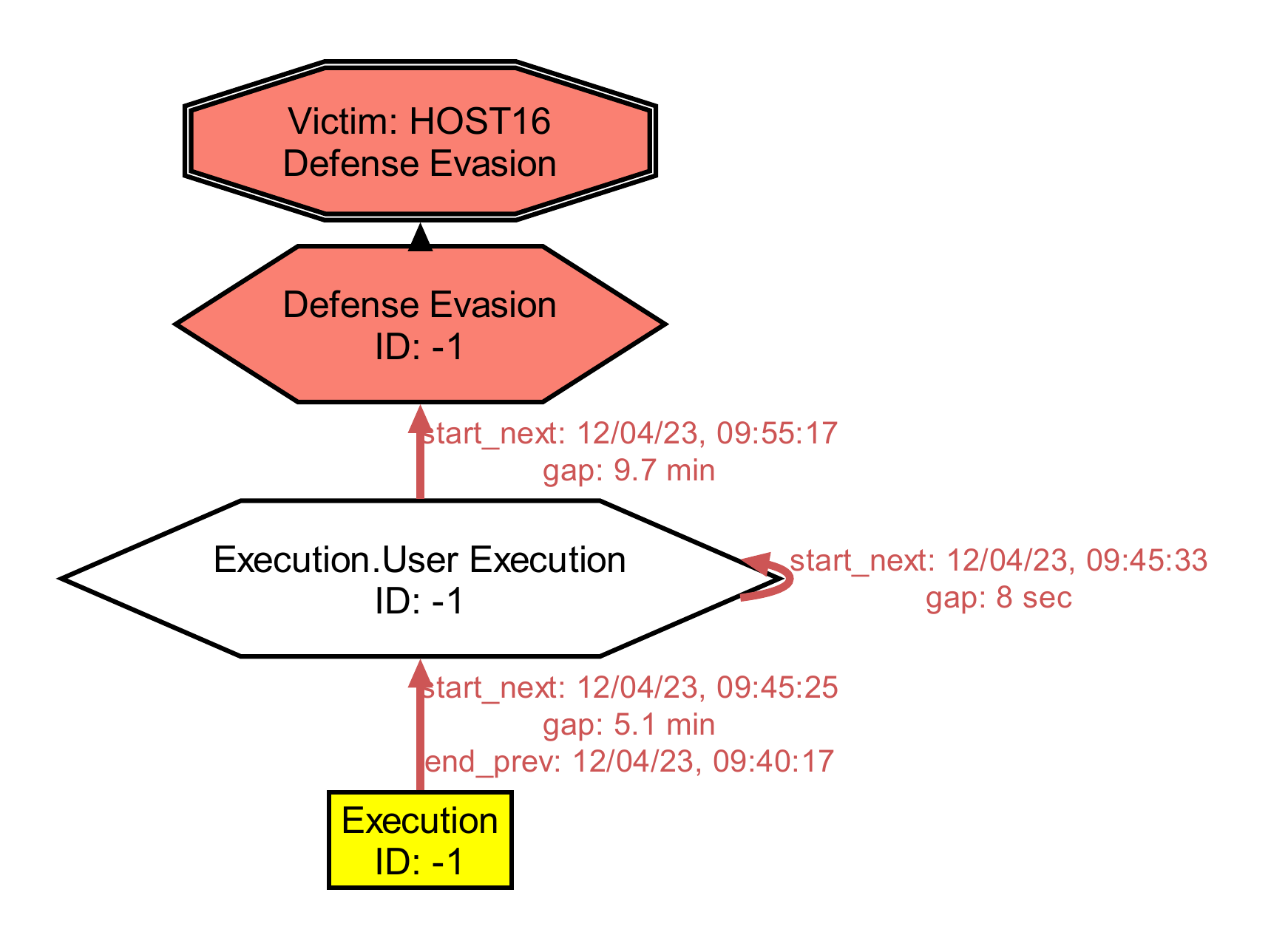}
    \label{sc1a}}
\subfloat[Scenario 1b (Evolving attack graphs)]{\includegraphics[width=0.31\linewidth]{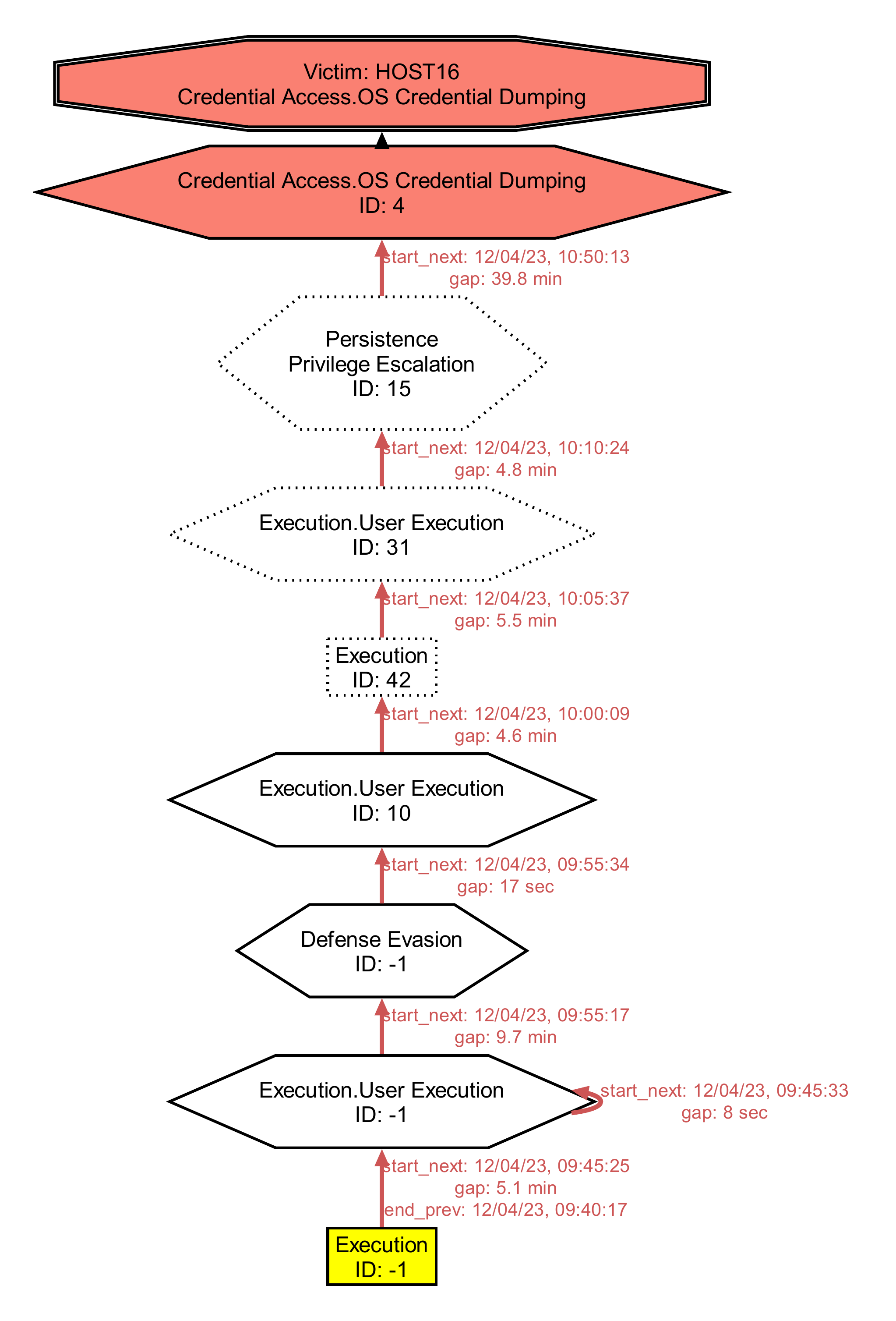}
\label{sc1b}
}
\subfloat[Scenario 2 (Action forecasting)]{\includegraphics[width=0.29\linewidth]{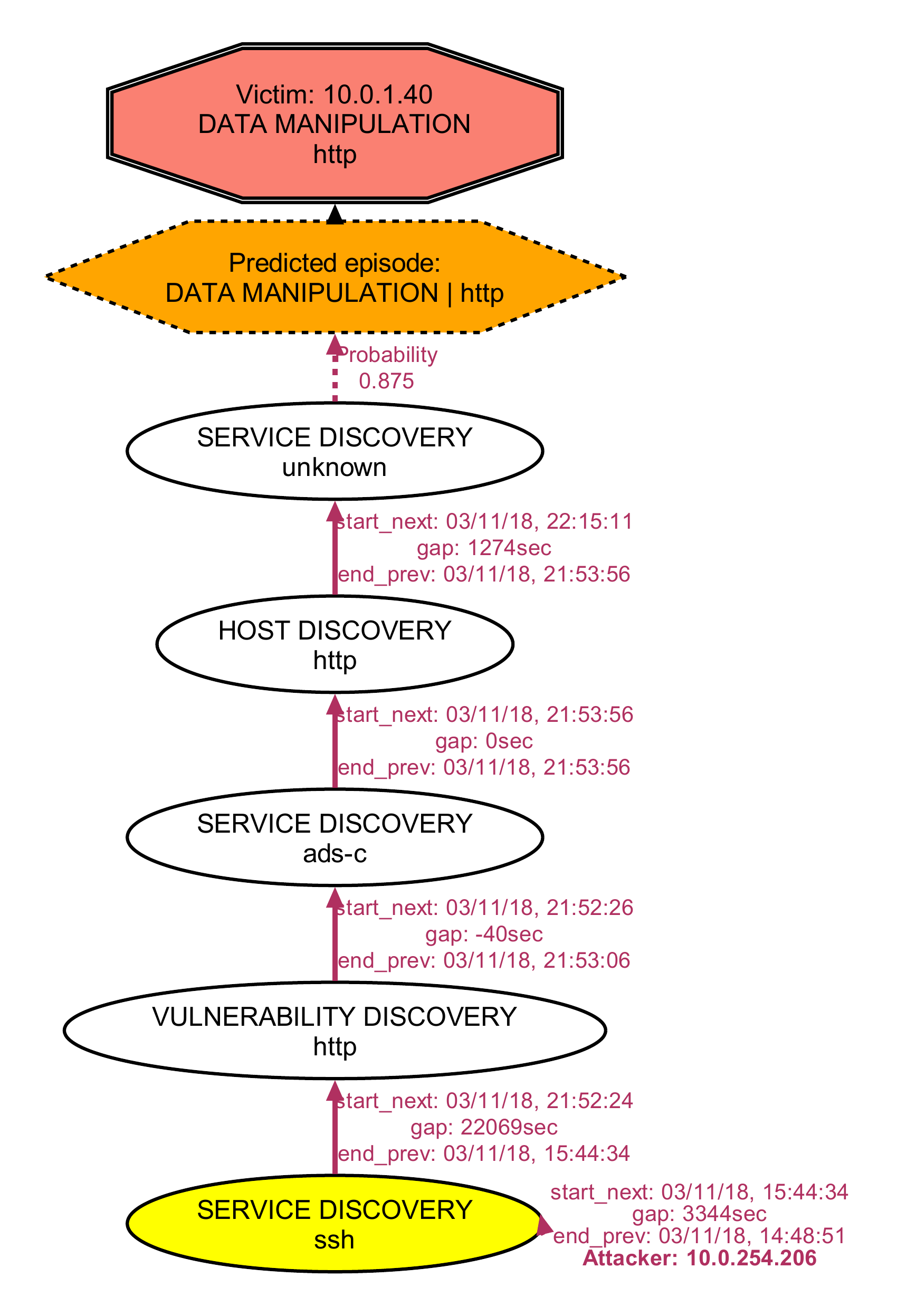}
\label{sc2}
}
\caption{Scenario 1 (a-b) shows two AGs generated from the Pentest dataset for Host16 approximately one hour apart from each other, representing an ongoing attack campaign. Scenario 2 (c) shows an AG generated from CPTC-2018 dataset for the host 10.0.1.40, suggesting that the attacker will likely perform data manipulation next using http with a probability of 87.5\%.}
    \label{fig:userstudy-scenarios}
\end{figure*}

\subsection{Evolving attack graphs}

We regenerate AGs every hour for the experimental datasets. 
For CPTC-2018,  we go from 24 AGs (averaging 6.6 vertices and 7.9 edges) in the first hour to 76 AGs (averaging 17.6 vertices and 35.8 edges) after 10 hours. 
For Pentest, we go from 11 AGs (averaging 3.9 vertices and 3.4 edges) in the first hour to 17 AGs (averaging 5.1 vertices and 5.3 edges) after 3 weeks. 
For Big environment, we go from 1 AG (2 vertices, 1 edge) in the first hour to 15 AGs (averaging 5.8 vertices and 6.5 edges) after 3 months.

Figures \ref{fig:userstudy-scenarios}(a-b) show two consecutive attack graphs generated for Host16 in Pentest, one hour apart from each other. Pentest contains EDR logs and is significantly sparser than CPTC-2018. In the first graph, we observe that a suspicious command was executed at 9:40, followed by a gap of 5 minutes after which a hack tool was executed. In the second graph, it becomes clear that it is an ongoing attack campaign, where more suspicious commands are executed, after which the hack tool is detected. At 10:10, privilege execution was performed, which resulted in credential dumping at 10:50. We utilize these AGs for our user study.


\subsection{User study: Attack graph usability}

\noindent{\textbf{Analyst workflow and challenges.}} 
The current analyst workflow typically revolves around investigating EDR logs, \eg Microsoft defender for endpoint (MDE) and ticketing systems like JIRA to collect relevant events, and manually correlate them with different sources to understand what is happening. For instance, P4 reports executing a JIRA search for a host under observation, and based on the results dig deeper into log portals in order to create a timeline for the host and correlate the alerts. 
Moreover, the participants do not utilize any specific visual analytics tools --- P2, P5, and P6 use the correlation graph shown by MDE, P5 additionally uses the Sentinel graph for rendering time charts, while P3 utilizes Azure Kusto queries to render time charts.

The participants report that the alert volume is one the biggest challenges they face in their workflows. P1 reports the difficulty of having to pull information from different sources to get a bigger picture of the threat landscape, while P4 reports the manual nature of this deep-dive investigation as time-consuming. \textit{Sometimes, an alert seemingly has a low severity, but when combined with the larger context, it appears to be a real threat.} 
This type of manual workload leads to alert fatigue, which is a well-studied phenomenon for SOC analysts \cite{fireeye}. All the study participants report either having experienced it themselves or know a colleague who experienced it. Specifically, P1 and P2 recall situations where the task repetition and alert volume caused them to miss nuances in critical alerts.

\noindent{\textbf{Scenario 1 (Evolving attack graphs):}} Figure \ref{sc1a} displays the attack graph we show to the participants and ask them to interpret it. All the participants were able to correctly reconstruct the attack story behind the AG by hovering over the vertices, exploring the alert signatures, and reading the timestamps. 
We then showed them an updated version of the AG, which was generated using additional alerts (see Figure \ref{sc1b}). The participants were able to identify the continuity between the two AGs. 
P5 believed that host16 was likely compromised since while a single OS credential dumping alert is considered to be a false positive (likely due to a misbehaving process), visualizing the entire attack chain that led to the OS credential dumping makes a pretty clear picture that something bigger is going on. 
P4 even believed that the situation is critical because of how much the AG had evolved from the previous one. \textit{Thus, the evolution of the AGs enables analysts to interpret the changing threat landscape in real-time and narrow down their investigation scope.} 

\noindent{\textbf{Scenario 2 (Action forecasting):}} We present Figure \ref{sc2} to the participants, showing a series of low-severity scanning episodes leading to the prediction of data manipulation with a probability of 87.5\% in CPTC-2018. Given only the attack path (without prediction), the participants believed that the attacker was likely trying to discover vulnerable hosts and services. Their unanimous advice was to scan the victim host, ensure that no unnecessary ports were open, and to make sure that the system was patched. 
They shared that the AG helped them identify the specific ports that needed to be checked, and that there were multiple scans in progress. 

Next, when given the knowledge about the prediction, the participants advised to isolate the victim host, and to investigate whether any sensitive data on the host had recently been modified. The urgency of their response seemed to be correlated to the prediction probability, \ie the higher the probability, the more likely they were to take action. Moreover, P5 also stated that they would keep the prediction in mind when investigating logs to look for additional related indicators.  

The participants shared that \textit{even if the prediction did not come true, it could be used to prioritize non-trivial attack paths.} For instance, most of the alerts were low-severity in Figure \ref{sc2}, but given the prediction, the participants would like to investigate what the prediction says, just to be safe. P4 stated: \textit{``[The prediction] gives you an idea of what could happen. Maybe you are getting an alert and you are like, yeah, that is not too bad, but then you do not realize that the next step in the graph would be something quite bad, which you want to prevent''.}

\noindent{\textbf{Comparison with existing tools:}} The participants reflected on the gaps filled by the alert-driven AGs that their current tools do not support.
The major theme was related to the volume of actionable information shown by such comprehensive AGs, which they have to manually discover and correlate in their current workflows. For instance, P2 suggested that the AGs are like MDE correlation graphs, but with much more information about time frames, and P6 highlighted the superiority of the AGs to JIRA queries, as JIRA cannot readily provide a timeline of the various attacker actions. 

The participants shared that the AGs were particularly useful for showing multiple (weak) signals that, together, form a serious threat. P4 appreciated the fact that the AGs link multiple alerts together by sharing: \textit{``having everything from [a] host in one place, that is an amazing feature.''} The visual representation of the sequence of events along with their timestamps helps to provide better insights into the threat. Alternatively, an analyst can easily lose track of the bigger picture when alerts from multiple hosts are triggered. P4 even wanted to view the AGs every time an alert was triggered just to check what other correlated alerts had been triggered in the past, and whether a similar attack path had been observed before. 


\section{Implications for incident response}

%
The S-PDFA model proposed in \cite{nadeem2021alert}  shows exceptional performance in discovering attack campaigns from significantly imbalanced alert datasets. 
We further modify the S-PDFA so it can forecast future actions given a partial attack path. 
This is helpful for creating evolving AGs that display increasing amounts of context as more alerts are introduced into the system. 
SAGE takes negligible time in learning the S-PDFA model, so it can be re-triggered as periodically as required. There is also no need to size-bound the AGs since the rSPDFA forecasts are based on at most $t=5$ symbols from the partial paths (although additional symbols might further improve the performance).
For the experiments, we create AGs from diverse datasets with vastly different distributions, and demonstrate that the proposed forecasting (HC) strategy achieves 67.27\% average prediction accuracy on one of the experimental datasets (a 57.17\% improvement over all baselines, on average). The results particularly demonstrate that the expectation maximization approach followed by the rSPDFA picks up non-trivial patterns, as there appears to be a non-linear relationship between the alert (severity) frequency and accuracy. Moreover, this approach also performs well in forecasting actions from previously unseen attack paths, thus showing promise for predicting zero-day attack campaigns. 

%
 %
%

The user study indicates a clear need for tooling that can aggregate alerts into attack paths, and also provide some kind of recommendation for next steps.
Note that we explicitly did not perform a comparative analysis against existing tools in the user study because none of the tools currently available on the market can derive attack paths and forecast future actions, as confirmed by the participants.
The participants were introduced to the alert-driven AGs only at the start of the user study, but they managed to correctly interpret the attacker actions and recommend remedial actions based on the forecasts, which speaks to the superiority of (relatively) simple visualizations over complex dashboards. 
We hope that these first results encourage further research into better and faster tooling for attack path analysis and prioritization.

\section{Related work}
Prior work on alert management has focused on alert correlation for clustering alerts likely belonging to the same attacker action~\cite{ning2002constructing,qin2004discovering,sadoddin2006alert,zhu2006alert,salah2013model,alserhani2016alert,wang2016alert,haas2018gac,shittu2015intrusion,mcelwee2017deep}. In this paper, however, we go beyond alert correlation by discovering attack paths. 
Therefore, we focus on two strains of research: dashboards for visualizing attack paths, and alert prediction for forecasting attacks.

\noindent\textbf{Attack visualization.} 
ASSERT \cite{yang2021near} builds so-called \textit{attacker behavior models} from streaming intrusion alerts. However, the dashboard visualizes statistical properties of the various ongoing attacks instead of attack paths. 
ATLAS \cite{alsaheel2021atlas} combines causality analysis and natural language processing to visualize attack campaigns derived from intrusion alerts. Given an alert, ATLAS predicts whether it was a part of a given attack story.  
Similarly, HeAT \cite{moskal2021heated} visualizes the sequence of alerts that led to a given critical alert. These methods reconstruct the past based on an alert of interest.

Attack graphs (AG) are popular for visualizing attack campaigns \cite{kaynar2016taxonomy}. They have historically been generated based on topological vulnerability analysis (TVA) \cite{noel2009advances}. MulVAL \cite{ou2005mulval} is a popular tool that creates AGs for a network topology given a list of pre-existing vulnerabilities. 
The TVA-based AGs however utilize information that is not readily available in operational settings. To overcome this limitation, Nadeem \etal developed SAGE \cite{nadeem2021alert}, the first machine learning based tool-chain that creates AGs directly from intrusion alerts \textit{without any expert input}. In this paper, we modify SAGE to create AGs in real-time that can forecast attacker actions.  

\noindent\textbf{Alert prediction.} Predicting the next likely attacker action is challenging because it requires efficient modeling algorithms for infrequent severe alerts. This is apparent from the very few studies that develop prediction systems to forecast attacker actions. 
%
%
%
Ramaki \etal \cite{ramaki2015real} predict the probability of the next likely attacker actions using Bayesian attack graphs, which are partially derived from network and vulnerability information, commonly unavailable in operational settings. 
Fava \etal \cite{fava2008projecting} and Thanthrige \etal \cite{thanthrige2016intrusion} utilize Markov models to predict future alerts. Particularly, Fava \etal utilize three Markov models to predict different alert characteristics. However, as shown in \cite{nadeem2021alert}, Markov models are not effective for context modeling, and utilizing more than one model negatively impacts the interpretability of the approach. Besides, we did not find open-source implementations for these works, and there is insufficient information in the papers to re-implement them.
In contrast, we propose a substantially simpler approach where a single automaton model is used for context modeling and action forecasting in a streaming setting. We release our code publicly.
%

%
%
%

%

\section{Conclusions}

%
We build a forecasting capability on top of alert-driven attack graphs (AGs) that lets us forecast the next likely attacker action given a partial path.
We also modify the framework to create AGs in real-time that evolve as new alerts are triggered. 
We empirically demonstrate that our proposed forecasting algorithm with a hybrid choice (HC) traversal strategy achieves an average top-3 accuracy of 67.27\% (a 57.17\% improvement over 3 baselines, on average). 
We also interview 6 SOC analysts to understand the efficacy of the proposed approach. 
The results show that the action forecasts help them take proactive remedial actions and prioritize critical incidents, while the evolving AGs enable them to maintain latest situational awareness.
We hope this work encourages further research into better tooling for attack path analysis \& prioritization. 

\paragraph{Acknowledgments.} The authors thank Sicco Verwer for his contributions to the rSPDFA algorithm.

\bibliographystyle{ACM-Reference-Format}
\bibliography{_main}

\appendix

\begin{figure}[t]
    \centering
    \includegraphics[width=0.8\linewidth]{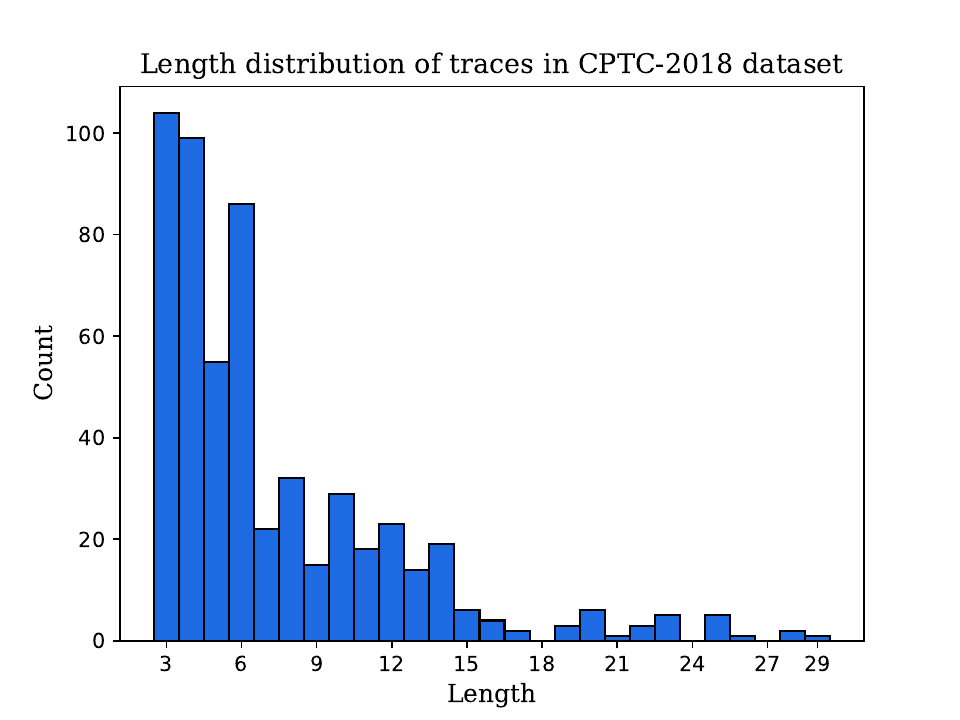}
    \caption{Length of traces in CPTC-2018. 344/555 (61.9\%) traces have 3 to 6 symbols}
    \label{pred:lendistrib}
\end{figure}

\begin{figure}[t]
    \centering
    \includegraphics[width=0.8\linewidth]{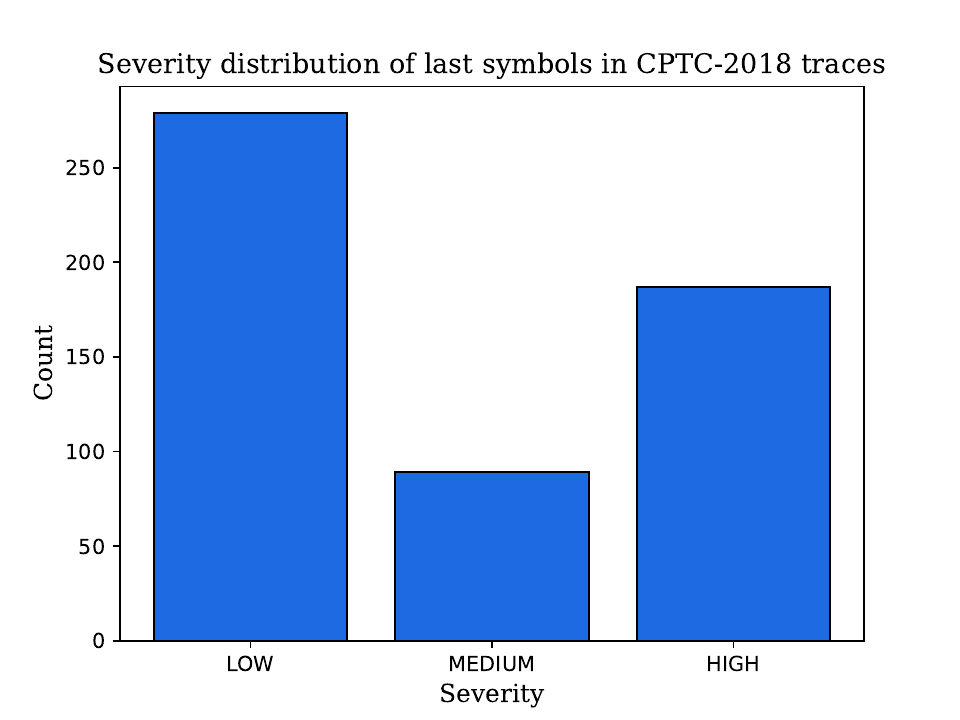}
    \caption{Severity distribution of the last symbols in CPTC-2018 traces. Most traces end in low-severity symbols.}
    \label{fig:sevdistrib}
\end{figure}

\begin{figure}[t]
    \centering
    \includegraphics[width=0.8\linewidth]{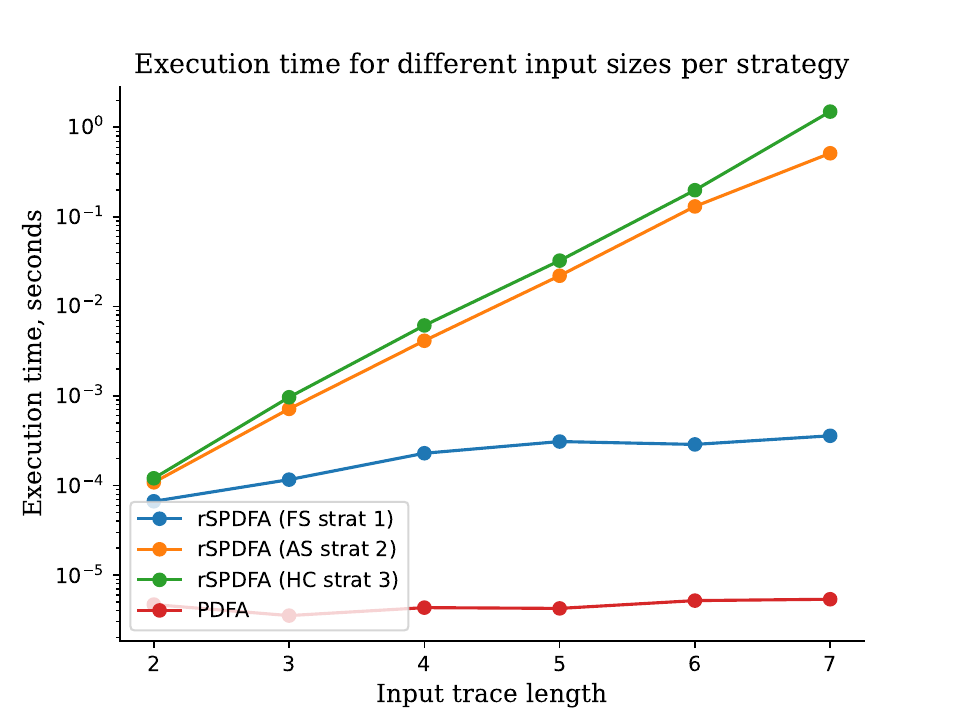}
    \caption{Execution time in seconds per input size (logarithmic scale). AS and HC are the most affected by the increase in trace length, as their execution time increases exponentially.}
    \label{fig:testing_time}
\end{figure}

\begin{figure}[t]
    \centering
    \includegraphics[width=0.8\linewidth]{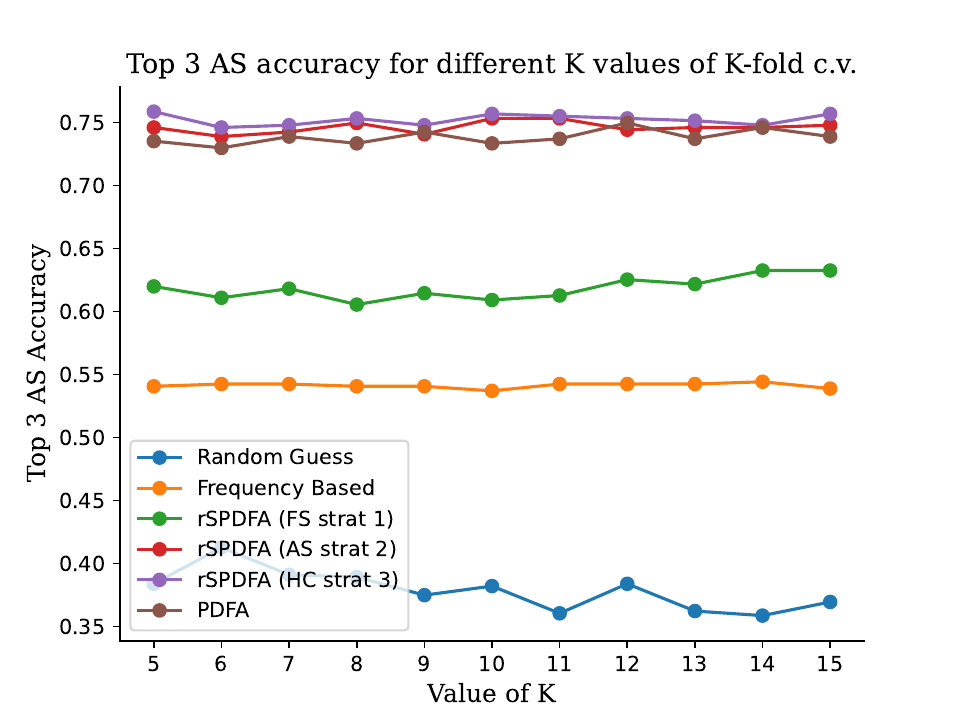}
    \caption{Accuracy of each strategy for a different K value used for K-fold cross-validation testing. We select \textit{k=5}.}
    \label{pred:kfold}
\end{figure}

\begin{figure}[t]
    \centering
    \includegraphics[width=0.8\linewidth]{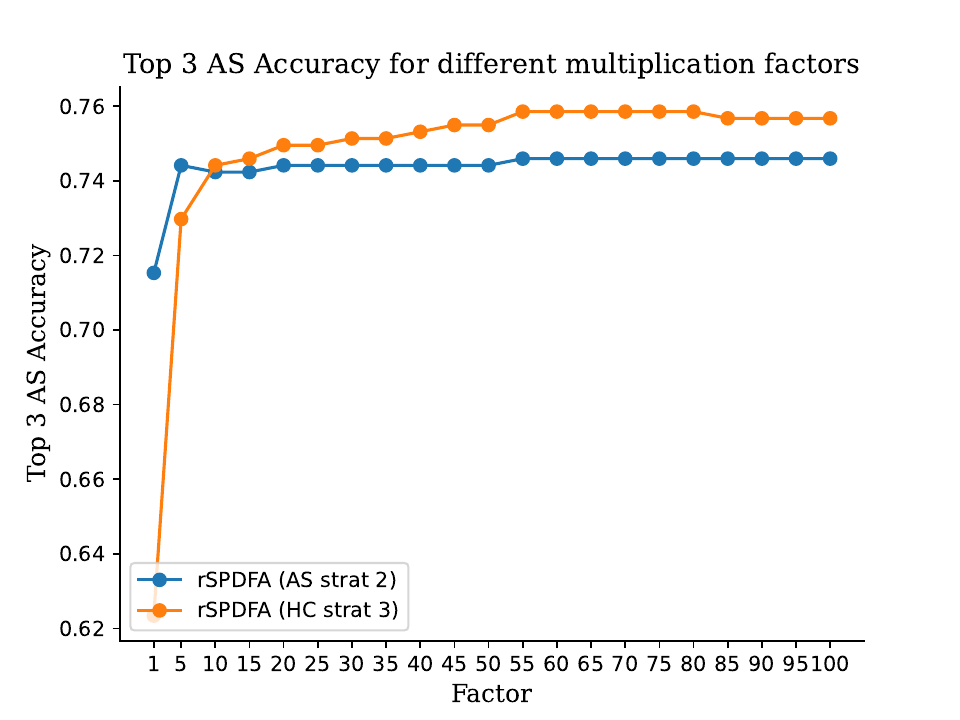}
    \caption{Accuracy for different multiplication factors $f$ for AS and HC. The accuracy increase plateaus at $55$ for both.}
    \label{fig:factor}
\end{figure}

\nop{
\begin{figure*}[t]
    \centering
\subfloat[]{\includegraphics[width=0.3\linewidth]{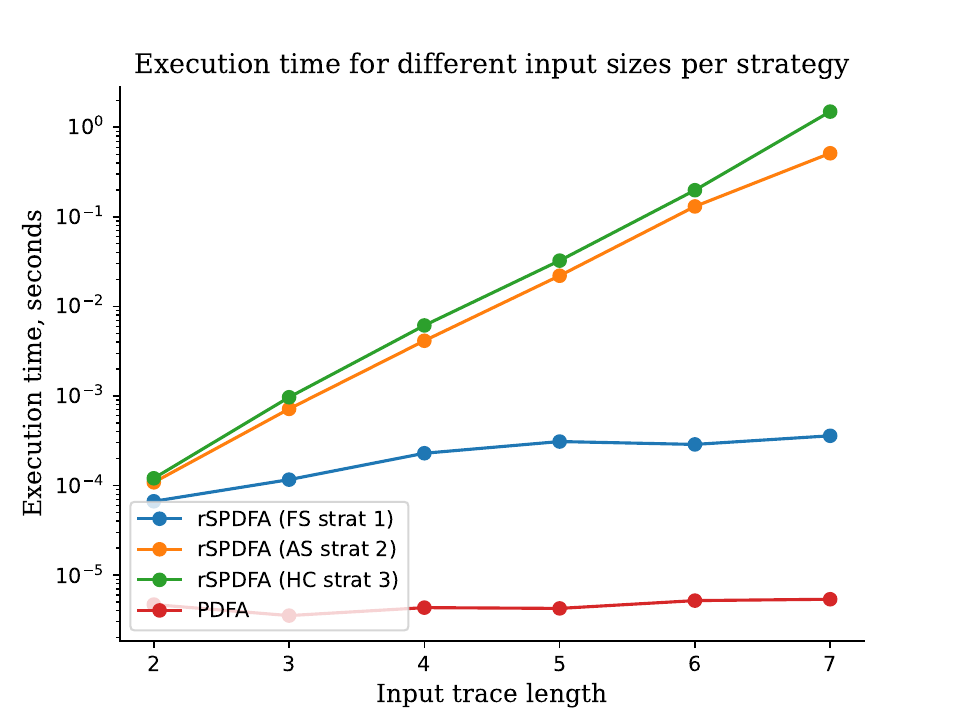}
\label{fig:testing_time}
}
\subfloat[]{\includegraphics[width=0.3\linewidth]{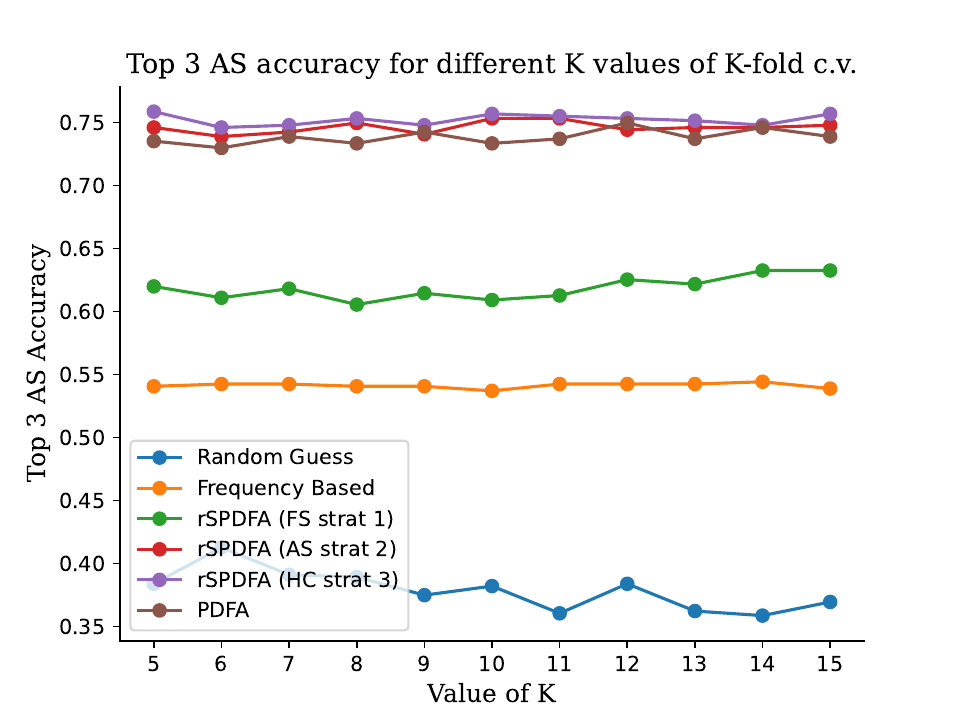}
\label{pred:kfold}
}
\subfloat[]{\includegraphics[width=0.3\linewidth]{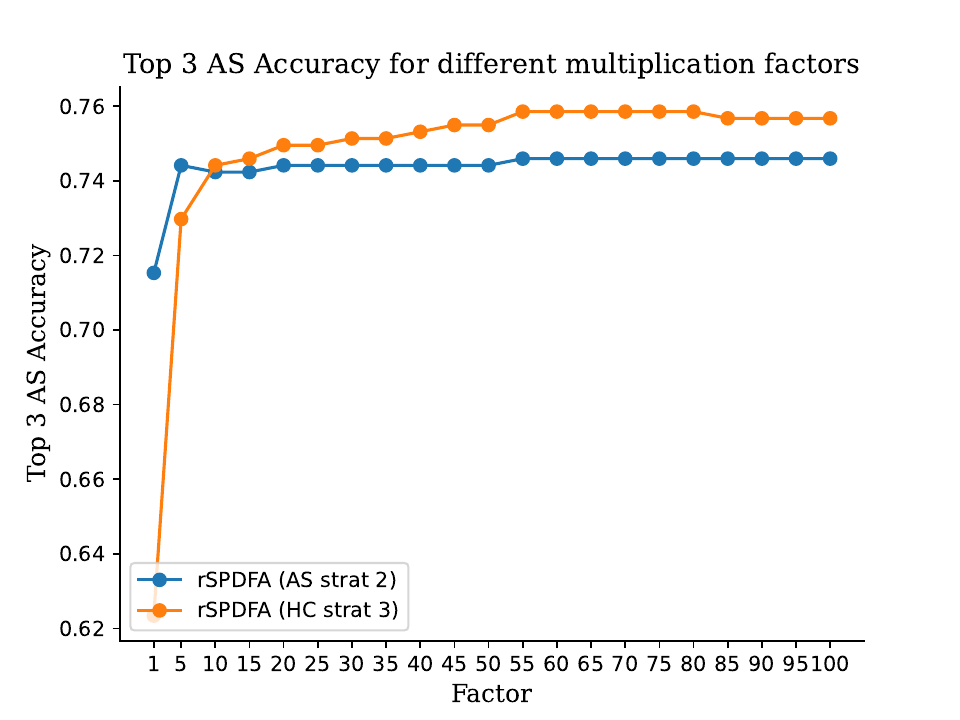}
\label{pred:factor}
}
\caption{(a): Execution time in seconds per different input sizes, plotted on a logarithmic scale. Strategies 2 and 3 are the most affected by the increase in trace length, as their execution time increases exponentially. (b): Accuracy of each strategy for a different K value used for K-fold cross-validation testing. We select \textit{k=5}. (c): Accuracy for different multiplication factors $f$ for strategies 2 and 3. The increase in accuracy plateaus at $
    55$ for both strategies.}
    \label{fig:overall}
\end{figure*}}

\section{Alert distribution CPTC-2018}

Figure \ref{pred:lendistrib} shows a histogram of the trace lengths in CPTC-2018. The trace lengths follow a typical long-tailed distribution, where most of the traces have 3-6 symbols. We, therefore, choose \textit{t=5} to be the length threshold to predict the 6th symbol. Moreover, Figure \ref{fig:sevdistrib} shows the make up of these traces. It shows that roughly half of the traces end in low-severity symbols and 18\% end in medium-severity symbols, likely capturing ongoing attack attempts. $\sim$36\% traces end in high severity symbols.

\section{Runtime evaluation}
We evaluate the impact of the trace length on the execution time of our action forecasting module. We learn the S-PDFA from the 555 CPTC-2018 traces. Subsequently, we execute each traversal strategy (FS, AS, HC) using input traces of varying lengths, from 2 to 7 symbols. Figure \ref{fig:testing_time} shows the runtime plots. We observe that the PDFA approach is the fastest because of determinism. Among the rSPDFA alternatives, FS is the fastest because it does not explore many paths, while AS and HC are among the slowest traversal strategies. 

\section{$K$-fold cross-validation}
 We test various values of k to find the optimal parameter for k-fold cross-validation. 
 We execute each method (baselines and our proposed strategies) with a varying k from range $\{5,15\}$ and compute the accuracy. Figure \ref{pred:kfold} shows that k does not have a significant impact on the accuracy of the best-performing strategy, partly because even smaller values of k capture a good overview of the testing set. The best accuracy is achieved at \textit{k=5}, which we choose as the parameter for the experiments.

\section{Multiplication factor $f$}
Strategies AS and HC are more flexible in terms of symbol matching. We want to prioritize full symbol matches over partial matches. We introduce a multiplication factor $f$ with each transition probability when computing the probability for a path. We test various values for $f$ to find the optimal parameter that maximizes accuracy. We compute the accuracy of each traversal strategy with a varying values of $f$ in range $\{1,95\}$. Figure \ref{fig:factor} shows the results. We can see that the biggest increase in accuracy happens at the beginning until the factor reaches a value of 10. Afterwards, a minor increase in accuracy is seen, which stops altogether at 55. By artificially inflating the transition probabilities, we prioritize paths with full symbol matches that lead to a higher prediction accuracy. We select $f=55$ for the experiments, as it is the first factor to obtain the best results for both strategies.

\balance

\end{document}